\documentclass[twocolumn,showpacs,preprintnumbers,amsmath,amssymb]{revtex4}

\usepackage{graphicx}
\usepackage{dcolumn}
\usepackage{bm}

\newcommand{\bq}{\begin{equation}}
\newcommand{\eq}{\end{equation}}
\newcommand{\bqa}{\begin{eqnarray}}
\newcommand{\eqa}{\end{eqnarray}}
\newcommand{\nn}{\nonumber \\}

\def\be     {\begin{equation}}
\def\ee     {\end{equation}}
\def\bea        {\begin{eqnarray}}
\def\eea        {\end{eqnarray}}
\def\bnn    {\begin{eqnarray*}}
\def\enn    {\end{eqnarray*}}

\begin{document}

\title{Nambu-Eliashberg theory for multi-scale quantum criticality : Application to ferromagnetic quantum criticality
in the surface of three dimensional topological insulators}
\author{Ki-Seok Kim and Tetsuya Takimoto}
\affiliation{Asia Pacific Center for Theoretical Physics, Hogil
Kim Memorial building 5th floor, POSTECH, Hyoja-dong, Namgu,
Pohang 790-784, Korea
\\ Department of Physics, Pohang University of Science and
Technology, Pohang, Gyeongbuk 790-784, Korea}
\date{\today}

\begin{abstract}
We develop an Eliashberg theory for multi-scale quantum
criticality, considering ferromagnetic quantum criticality in the
surface of three dimensional topological insulators. Although an
analysis based on the random phase approximation has been
performed for multi-scale quantum criticality, an extension to an
Eliashberg framework was claimed to be far from triviality in
respect that the self-energy correction beyond the random phase
approximation, which originates from scattering with $z = 3$
longitudinal fluctuations, changes the dynamical exponent $z = 2$
in the transverse mode, explicitly demonstrated in nematic quantum
criticality. A novel ingredient of the present study is to
introduce an anomalous self-energy associated with the spin-flip
channel. Such an anomalous self-energy turns out to be essential
for self-consistency of the Eliashberg framework in the
multi-scale quantum critical point because this off diagonal
self-energy cancels the normal self-energy exactly in the low
energy limit, preserving the dynamics of both $z = 3$ longitudinal
and $z = 2$ transverse modes. This multi-scale quantum criticality
is consistent with that in a perturbative analysis for the nematic
quantum critical point, where a vertex correction in the fermion
bubble diagram cancels a singular contribution due to the
self-energy correction, maintaining the $z = 2$ transverse mode.
We also claim that this off diagonal self-energy gives rise to an
artificial electric field in the energy-momentum space in addition
to the Berry curvature. We discuss the role of such an anomalous
self-energy in the anomalous Hall conductivity.
\end{abstract}


\maketitle

\section{Introduction}

Investigation on the role of the momentum-space Berry curvature
has driven an interesting direction of research in modern
condensed matter physics, covering weak anti-localization,
anomalous charge and spin Hall effects, and topological insulators
\cite{Review_Berry,Review_AHE,Review_TI}. A recent trend is to
study interplay between the topological band structure and
interaction, suggesting that topological excitations such as
skyrmions and vortices carry fermion quantum numbers of electric
charge or spin via the associated topological term
\cite{Soliton_TextBook}.

Quantum criticality has also driven an important direction of
research in modern condensed matter physics, particulary focusing
on the nature of non-Fermi liquid physics and mechanism of
superconductivity out of the non-Fermi liquid state
\cite{Review_QCP}. Combined with the topological band structure,
critical degrees of freedom given by topological excitations in
the duality picture will carry fermion quantum numbers, thus a
novel type of non-Fermi liquid physics may arise.

In this paper we study ferromagnetic quantum criticality in the
surface of three dimensional topological insulators from the side
of a disordered phase. A recent analysis based on the random phase
approximation (RPA) has shown a multi-scale quantum critical
point, where longitudinal spin fluctuations are described by the
dynamical critical exponent $z = 3$ while transverse modes are
given by $z = 2$ \cite{Xu_RPA}. Although scaling can be
demonstrated within this analysis, stability of the RPA dynamics
is not guaranteed beyond this approximation. In order to describe
scaling near quantum criticality, one should go beyond the
perturbative analysis, introducing quantum corrections fully
self-consistently. An Eliashberg theory is desirable, where
self-energy corrections are taken into account fully
self-consistently in the one loop level \cite{Rech_FMQCP}. The
scaling expression of the free energy can be derived within the
Eliashberg approximation, where critical exponents are determined
\cite{Kim_LW}. Unfortunately, an extension to an Eliashberg
framework was claimed to be far from triviality in multi-scale
quantum criticality because the self-energy correction, which
originates from scattering with $z = 3$ critical fluctuations,
turns out to change the dynamical exponent $z = 2$ in the
transverse mode, explicitly demonstrated in nematic quantum
criticality \cite{Garst_Namatic_QCP}.

In this study we construct an Eliashberg theory for multi-scale
quantum criticality, considering ferromagnetic quantum criticality
in the topological surface. A novel ingredient is to introduce an
anomalous self-energy associated with the spin-flip channel. Such
an anomalous self-energy turns out to be essential for
self-consistency of the Eliashberg framework because this off
diagonal self-energy cancels the normal self-energy exactly in the
low energy limit, preserving the dynamics of both $z = 3$ and $z =
2$ critical modes. This multi-scale quantum criticality is
consistent with that in a perturbative analysis of the nematic
quantum critical point, where a vertex correction in the fermion
bubble diagram cancels a singular contribution due to the
self-energy correction, maintaining $z = 2$ for the transverse
mode \cite{Garst_Namatic_QCP}.
%
%
We claim that such an off-diagonal self-energy generates an
artificial electric field in the energy-momentum space
\cite{Shindou_E} in addition to the Berry curvature. Based on the
semi-classical approximation for the Hall conductivity, we find an
anomalous divergent behavior at zero temperature. We discuss
physical implication of this phenomenology in connection with the
topological nature of Dirac fermions.
%
%

\section{Hertz-Moriya-Millis theory in the surface of the
topological insulator}

\subsection{Model}

We start from an effective field theory for ferromagnetic quantum
criticality in the surface of topological insulators \bqa && Z =
\int D \boldsymbol{\psi}_{\alpha} D \boldsymbol{\phi}_{i} e^{-
\int_{0}^{\beta} d \tau \int d^{2} \boldsymbol{r} {\cal L} } , \nn
&& {\cal L} = \boldsymbol{\psi}_{\alpha}^{\dagger}
[(\partial_{\tau} - \mu) \delta_{\alpha\alpha'} + i v_{f}
\boldsymbol{\hat{z}} \cdot (\boldsymbol{\sigma} \times
\boldsymbol{\nabla})_{\alpha\alpha'} ] \boldsymbol{\psi}_{\alpha'}
\nn && + |\partial_{\tau} \boldsymbol{\phi}_{i}|^{2} + v_{b}^{2}
|\boldsymbol{\nabla} \boldsymbol{\phi}_{i}|^{2} + m_{b}^{2}
|\boldsymbol{\phi}_{i}|^{2} + V(|\boldsymbol{\phi}_{i}|) \nn &&  +
g \boldsymbol{\phi}_{i} \boldsymbol{\psi}_{\alpha}^{\dagger}
\boldsymbol{\sigma}_{\alpha\alpha'}^{i}
\boldsymbol{\psi}_{\alpha'} , \eqa where $\boldsymbol{\psi} =
\left( \begin{array} {c} \psi_{\uparrow} \\
\psi_{\downarrow} \end{array} \right)$ represents an electron
field in the Nambu-spinor representation and $\boldsymbol{\phi} =
\phi_{x} \boldsymbol{\hat{x}} + \phi_{y} \boldsymbol{\hat{y}}$
corresponds to a spin excitation. We consider the case of an XY
spin for simplicity, which can be generalized into the case of
O(3) spin. $v_{f}$ is the Dirac velocity and $\mu$ is the chemical
potential for surface electrons. $\boldsymbol{\sigma}$ is the
Pauli matrix, associated with the spin state. $v_{b}$ is the
spin-fluctuation velocity and $m_{b}^{2}$ is the inverse of
correlation-length for spin fluctuations, corresponding to an XY
ordered state when $m_{b}^{2} < 0$ and a quantum disordered phase
when $m_{b}^{2} > 0$. $V(|\boldsymbol{\phi}_{i}|)$ is an effective
potential for spin fluctuations, where an easy plane anisotropy
allows the XY spin dynamics only. $g$ is an effective coupling
constant between electrons and spin excitations, where tuning $g$
leads the spin correlation-length to diverge, resulting in quantum
criticality. The quantum critical point between the XY ordered and
disordered phases is defined by the vanishing effective mass for
spin fluctuations, tuned by the coupling parameter $g$.

This effective field theory would be realized when an insulating
ferromagnet lies on the surface of topological insulators. In this
case ferromagnetic spin fluctuations denoted by
$\boldsymbol{\phi}_{i}$ describe those in the insulating
ferromagnet. Then, the coupling parameter $g$ represents the
strength of couplings between spins in the ferromagnet and
itinerant electrons on the topological surface. Another similar
setting is that ferromagnetism will be induced by doped magnetic
impurities. It was demonstrated that Ruderman-Kittel-Kasuya-Yosida
interactions between doped magnetic impurities are ferromagnetic
\cite{FM_RKKY}. Through this exchange mechanism, magnetic atoms
are expected to form a ferromagnetically ordered film, deposited
uniformly on the surface of a topological insulator. In this case
a possible ferromagnetic transition may be realized, controlling
the distance of doped impurities in the regular impurity pattern.

We see two important features in dynamics of electrons on the
topological surface state. The first is that spin dynamics is
quenched to the orbital motion of surface electrons. As a result,
one finds that the role of an external magnetic field differs from
that in graphene, where the pseudo-spin quantum number is
constrained to the orbital motion. In particular, one of the
authors could reveal that the Kondo effect and Friedel oscillation
around the magnetic impurity should be modified in the presence of
an external magnetic field, compared with the graphene structure
\cite{Tran_Kim_MI_TI}. The second aspect is more fundamental, that
is, only an odd number of Dirac fermions is allowed to appear in
the topological surface state \cite{Review_TI}. As a result,
quantum anomaly is realized, giving rise to nontrivial topological
properties of the system \cite{Ryu_Dirac_Mass}. For example, a
vortex in the XY ordered state will carry an electric charge,
giving rise to an anomalous Hall effect, different from the
"conventional" anomalous Hall effect in conventional ferromagnets
\cite{Review_AHE}. It should be noted that this phenomenology does
not occur in usual condensed matter systems because the quantum
anomaly is often cancelled by the so called fermion doubling
effect \cite{Review_TI}.

Recently, interaction effects have been investigated in the
graphene structure, proposing an interesting phase diagram based
on the quantum Monte Carlo simulation \cite{Graphene_QMC}.
Immediately, such a phase diagram was interpreted in various
analytical scenarios, where the singlet-channel interaction was
emphasized \cite{Graphene_Interaction}. On the other hand, we
focus on the role of spin-flip scattering, allowed by interactions
in the spin-triplet channel. We introduce an anomalous electron
self-energy associated with the spin-flip scattering, and discuss
its possible implication in connection with the topological nature
of the surface state.

\subsection{Eliashberg approximation}

One can perform an Eliashberg approximation for the effective
Lagrangian [Eq. (1)] at the quantum critical point ($m_{b}^{2} =
0$), where both fermion and boson self-energy corrections are
determined fully self-consistently in the one-loop level.
Considering the cumulant expansion up to the second order, we can
find an effective action
\begin{widetext}
\bqa && {\cal S}_{QC} = \int d^{3} \boldsymbol{x} \int d^{3}
\boldsymbol{x}' \Bigl(
\boldsymbol{\psi}_{\alpha}^{\dagger}(\boldsymbol{x})
[\{(\partial_{\tau} - \mu) \delta_{\alpha\alpha'} + i v_{f}
\boldsymbol{\hat{z}} \cdot (\boldsymbol{\sigma} \times
\boldsymbol{\nabla})_{\alpha\alpha'}\}
\delta(\boldsymbol{x}-\boldsymbol{x}') +
\boldsymbol{\Sigma}_{\alpha\alpha'}(\boldsymbol{x}-\boldsymbol{x}')
] \boldsymbol{\psi}_{\alpha'}(\boldsymbol{x}') \nn && - N_{\sigma}
\boldsymbol{\Sigma}_{\alpha\alpha'}(\boldsymbol{x}-\boldsymbol{x}')
\boldsymbol{G}_{\alpha'\alpha}(\boldsymbol{x}'-\boldsymbol{x})
\Bigr) \nn && + \int d^{3} \boldsymbol{x} \int d^{3}
\boldsymbol{x}' \Bigl( \boldsymbol{\phi}_{i}(\boldsymbol{x})
[(-\partial_{\tau}^{2} - v_{b}^{2} \boldsymbol{\nabla}^{2})
\delta(\boldsymbol{x}-\boldsymbol{x}') \delta_{ij} +
\boldsymbol{\Pi}_{ij} (\boldsymbol{x}-\boldsymbol{x}')]
\boldsymbol{\phi}_{j}(\boldsymbol{x}') - \boldsymbol{\Pi}_{ij}
(\boldsymbol{x}-\boldsymbol{x}') \boldsymbol{D}_{ij}
(\boldsymbol{x}-\boldsymbol{x}') \Bigr) \nn && + \frac{g^{2}}{2}
\int d^{3} \boldsymbol{x} \int d^{3} \boldsymbol{x}'
\boldsymbol{D}_{ij} (\boldsymbol{x}-\boldsymbol{x}')
\boldsymbol{tr} \Bigl( \boldsymbol{\sigma}_{i}
\boldsymbol{G}(\boldsymbol{x}-\boldsymbol{x}')
\boldsymbol{\sigma}_{j}
\boldsymbol{G}(\boldsymbol{x}'-\boldsymbol{x}) \Bigr) . \eqa
\end{widetext}
$\boldsymbol{\Sigma}_{\alpha\alpha'}(\boldsymbol{x}-\boldsymbol{x}')$
and
$\boldsymbol{G}_{\alpha'\alpha}(\boldsymbol{x}'-\boldsymbol{x})$
are the electron self-energy and Green's function, respectively,
where $\alpha$ and $\alpha'$ represent spin states, $\uparrow$ and
$\downarrow$. $\boldsymbol{\Pi}_{ij}
(\boldsymbol{x}-\boldsymbol{x}')$ and $\boldsymbol{D}_{ij}
(\boldsymbol{x}-\boldsymbol{x}')$ are the spin-fluctuation
self-energy and Green's function, respectively, where $i$ and $j$
express spin directions, $x$ and $y$. All repeated indices are
summed. The last term is called the Luttinger-Ward functional
\cite{LW}, which is the key ingredient for an interaction effect
obtained in the Eliashberg approximation.

Integrating over electrons and spin fluctuations, we obtain an
effective free energy as a functional for self-energies
\begin{widetext} \bqa &&
F_{LW}[\boldsymbol{\Sigma}(i\omega),\boldsymbol{\Pi}(\boldsymbol{q},i\Omega)]
= - \frac{N_{\sigma}}{\beta} \sum_{i\omega} \int \frac{d^{2}
\boldsymbol{k}}{(2\pi)^{2}} \boldsymbol{tr}_{\alpha\alpha'}
\Bigl\{ \ln [- \boldsymbol{G}^{-1}(\boldsymbol{k},i\omega)] +
\boldsymbol{\Sigma}(\boldsymbol{k},i\omega)
\boldsymbol{G}(\boldsymbol{k},i\omega) \Bigr\} \nn && +
\frac{1}{\beta} \sum_{i\Omega} \int \frac{d^{2}
\boldsymbol{q}}{(2\pi)^{2}} \boldsymbol{tr}_{ij} \Bigl\{ \ln [
\boldsymbol{D}^{-1}(\boldsymbol{q},i\Omega)] -
\boldsymbol{\Pi}(\boldsymbol{q},i\Omega)
\boldsymbol{D}(\boldsymbol{q},i\Omega) \Bigr\} \nn && + N_{\sigma}
\frac{g^{2}}{2} \frac{1}{\beta} \sum_{i\Omega} \int \frac{d^{2}
\boldsymbol{q}}{(2\pi)^{2}} \frac{1}{\beta} \sum_{i\omega} \int
\frac{d^{2} \boldsymbol{k}}{(2\pi)^{2}}
\boldsymbol{D}_{ij}(\boldsymbol{q},i\Omega)
\boldsymbol{tr}_{\alpha\alpha'} \Bigl( \boldsymbol{\sigma}_{i}
\boldsymbol{G}(\boldsymbol{k}+\boldsymbol{q},i\omega+i\Omega)
\boldsymbol{\sigma}_{j} \boldsymbol{G}(\boldsymbol{k},i\omega)
\Bigr) , \eqa
\end{widetext}
where \bqa && \boldsymbol{G}(\boldsymbol{k},i\omega) = \Bigl(
\boldsymbol{g}^{-1} (\boldsymbol{k},i\omega) -
\boldsymbol{\Sigma}(\boldsymbol{k},i\omega) \Bigr)^{-1} , \nn &&
\boldsymbol{D}(\boldsymbol{q},i\Omega) = \Bigl( [\Omega^{2} +
v_{b}^{2} \boldsymbol{q}^{2}] \boldsymbol{I} +
\boldsymbol{\Pi}(\boldsymbol{q},i\Omega) \Bigr)^{-1} \eqa are
fully renormalized propagators of electrons and spin fluctuations,
respectively. \bqa && \boldsymbol{g} (\boldsymbol{k},i\omega) =
\frac{(i \omega + \mu ) \boldsymbol{I} + v_{f} \epsilon_{ij}
\boldsymbol{k}_{i} \boldsymbol{\sigma}_{j}}{(i \omega + \mu )^{2}
- v_{f}^{2} \boldsymbol{k}^{2} } \nonumber \eqa is the bare
propagator of electrons, where $\epsilon_{ij}$ is the two
dimensional antisymmetric tensor with $i, j = x, y$. $N_{\sigma}$
represents the spin degeneracy, in our case $N_{\sigma} = 2$.
$\boldsymbol{tr}_{\alpha\alpha'}$ and $\boldsymbol{tr}_{ij}$ mean
trace for spin states and spin directions, respectively.

Minimizing the free energy functional with respect to both
self-energy corrections, we obtain fully self-consistent coupled
Eliashberg equations for electron and boson self-energies
\begin{widetext}
\bqa && \boldsymbol{\Pi}_{ij}(\boldsymbol{q},i\Omega) = N_{\sigma}
\frac{g^{2}}{2} \frac{1}{\beta} \sum_{i\omega} \int \frac{d^{2}
\boldsymbol{k}}{(2\pi)^{2}} \boldsymbol{tr}_{\alpha\alpha'} \Bigl(
\boldsymbol{\sigma}_{i}
\boldsymbol{G}(\boldsymbol{k}+\boldsymbol{q},i\omega+i\Omega)
\boldsymbol{\sigma}_{j} \boldsymbol{G}(\boldsymbol{k},i\omega)
\Bigr) , \nn && \boldsymbol{\Sigma}(\boldsymbol{k},i\omega) =
g^{2} \frac{1}{\beta} \sum_{i\Omega} \int \frac{d^{2}
\boldsymbol{q}}{(2\pi)^{2}}
\boldsymbol{D}_{ij}(\boldsymbol{q},i\Omega)
\boldsymbol{\sigma}_{i}
\boldsymbol{G}(\boldsymbol{k}+\boldsymbol{q},i\omega+i\Omega)
\boldsymbol{\sigma}_{j} . \eqa
\end{widetext}
As discussed before, the triplet interaction channel gives rise to
the spin-flip scattering, corresponding to an off-diagonal term in
the electron self-energy matrix. The normal Eliashberg self-energy
is proposed to depend on frequency only because the momentum
dependence is regular and the singular behavior can be shown from
the frequency dependence \cite{Rech_FMQCP,Rech_FMQCP_Comment}. On
the other hand, the momentum dependence for the anomalous
self-energy has not been clarified yet.
%
%
We propose the following ansatz for the electron self-energy \bqa
&& \boldsymbol{\Sigma}(\boldsymbol{k},i\omega) = \Sigma(i\omega)
\boldsymbol{I} + \Phi(i\omega) \epsilon_{ij}
\boldsymbol{\hat{k}}_{i} \boldsymbol{\sigma}_{j} , \eqa where the
momentum dependence for the off-diagonal self-energy is given by
the same dependence as the topological band structure. In other
words, the off-diagonal self-energy is added to the dispersion
relation as the normal self-energy in the conventional system.
$\boldsymbol{\hat{k}} = \boldsymbol{k}/|\boldsymbol{k}|$ is the
unit vector.

\subsection{Polarization function}

Inserting the self-energy expression into the Green's function, we
obtain \bqa && \boldsymbol{G}(\boldsymbol{k},i\omega) = \frac{[i
\omega + \mu - \Sigma(i\omega)] \boldsymbol{I} + [v_{f}
|\boldsymbol{k}| + \Phi(i\omega) ]\epsilon_{ij}
\boldsymbol{\hat{k}}_{i} \boldsymbol{\sigma}_{j}}{[i \omega + \mu
- \Sigma(i\omega)]^{2} - [v_{f} |\boldsymbol{k}| + \Phi(i\omega)
]^{2} } . \nonumber \eqa Based on this expression, we can evaluate
the boson self-energy, given by the fermion polarization bubble. A
detailed procedure is shown in appendix A.

It turns out that spin dynamics at the XY ferromagnetic quantum
critical point is given by the following effective Lagrangian
\begin{widetext}
\bqa && {\cal L}_{\phi} = \Bigl( {\cal C}_{L} g^{2}
\frac{|\Omega|}{v_{f} |\boldsymbol{q}|} + v_{L}^{2}
|\boldsymbol{q}|^{2} \Bigr)
\boldsymbol{\phi}_{i}(\boldsymbol{q},i\Omega) \boldsymbol{P}_{ij}
\boldsymbol{\phi}_{j}(-\boldsymbol{q},-i\Omega) + \Bigl( {\cal
C}_{T} g^{2} \frac{\Omega^{2}}{v_{f}^{2} |\boldsymbol{q}|^{2}} +
v_{T}^{2} |\boldsymbol{q}|^{2} \Bigr)
\boldsymbol{\phi}_{i}(\boldsymbol{q},i\Omega) (\delta_{ij} -
\boldsymbol{P}_{ij})
\boldsymbol{\phi}_{j}(-\boldsymbol{q},-i\Omega) , \eqa
\end{widetext}
where \bqa && \boldsymbol{P}_{ij} =
\frac{\boldsymbol{q}_{i}\boldsymbol{q}_{j}}{|\boldsymbol{q}|^{2}}
\nonumber \eqa is the projection operator to satisfy
$\boldsymbol{P}_{ik} \boldsymbol{P}_{kj} = \boldsymbol{P}_{ij}$.
${\cal C}_{L(T)}$ is a constant of the order of $1$ and $v_{L(T)}$
is the renormalized velocity for spin fluctuations, where $L(T)$
represents the longitudinal (transverse) mode. An important
feature for spin dynamics is multi-scale quantum criticality,
where the longitudinal spin dynamics is given by the dynamical
exponent $z = 3$ while the transverse one is described by $z = 2$.
The multi-scale quantum criticality can be interpreted as follows.
The longitudinal spin dynamics is not involved with the spin-flip
scattering, described by the $z = 3$ ferromagnetic quantum
criticality. On the other hand, the transverse dynamics is
associated with the spin-flip process. In the topological surface
state this process appears with finite momentum transfer, similar
to the process with antiferromagnetic fluctuations. As a result,
the transverse spin dynamics is described by $z = 2$. We note that
the self-consistent Eliashberg approximation gives essentially the
same result as the random phase approximation \cite{Xu_RPA}.

\subsection{Electron self-energy}

The spin-fluctuation propagator is given by \bqa &&
\boldsymbol{D}_{ij}(\boldsymbol{q},i\Omega) = \frac{
\boldsymbol{P}_{ij}}{{\cal C}_{L} g^{2} \frac{|\Omega|}{v_{f}
|\boldsymbol{q}|} + v_{L}^{2} |\boldsymbol{q}|^{2}} +
\frac{\delta_{ij} - \boldsymbol{P}_{ij}}{{\cal C}_{T} g^{2}
\frac{\Omega^{2}}{v_{f}^{2} |\boldsymbol{q}|^{2}} + v_{T}^{2}
|\boldsymbol{q}|^{2}} \nn && \equiv D_{L}(\boldsymbol{q},i\Omega)
\boldsymbol{P}_{ij} + D_{T}(\boldsymbol{q},i\Omega) (\delta_{ij} -
\boldsymbol{P}_{ij}) , \eqa where
$D_{L(T)}(\boldsymbol{q},i\Omega)$ is the kernel of the
longitudinal (transverse) mode. Based on this spin-fluctuation
propagator, we find an electron self-energy in the Eliashberg
approximation. An important point is the role of the multi-scale
quantum criticality in the electron self-energy. It turns out that
the electron self-energy is governed by the longitudinal spin
dynamics at low energies because the $z = 3$ critical dynamics
gives rise to more singular corrections than the $z = 2$
transverse mode. Actually, this result was expected in the
Eliashberg approximation. But, it was claimed that a more
complicated structure is hidden beyond the Eliashberg
approximation \cite{Garst_Nematic}, which will be discussed in the
last section.

Our novel point is that the anomalous self-energy correction is
exactly the same as the normal one, given by
\begin{widetext} \bqa && \Sigma(i\omega) = -
\Phi(i\omega) \approx - i \frac{g^{2}}{2} \frac{1}{\beta}
\sum_{i\Omega} \int_{0}^{\infty} d q q  \Bigl(
D_{L}(\boldsymbol{q},i\Omega) + D_{T}(\boldsymbol{q},i\Omega)
\Bigr) \frac{ \mbox{sgn}(\omega+\Omega) }{
\sqrt{(\omega+\Omega)^{2} + v_{f} q^{2}}}  \propto |\omega|^{2/3}
, \eqa
\end{widetext}
where the last approximate equality is given by the $z = 3$
longitudinal mode $D_{L}(\boldsymbol{q},i\Omega)$. This
equivalence is at the heart of the self-consistency in the
Nambu-Eliashberg framework. As discussed in the introduction, the
self-energy correction in the fermion bubble diagram changes the
$z = 2$ dynamical exponent in the transverse mode. Actually, one
can see this effect from Eqs. (A9) and (A13) in appendix A. If the
anomalous self-energy is not introduced, the normal self-energy
contribution dominates over the bare frequency, modifying the
dynamical exponent of the transverse mode from $z = 2$ to $z =
12/5$. However, introduction of the anomalous self-energy cancels
the normal self-energy exactly in the low energy limit, recovering
the RPA result \cite{Xu_RPA}.

It is interesting to compare the mechanism of cancellation with
the perturbative analysis in nematic quantum criticality, where
next leading corrections given by electron self-energy and
Maki-Thompson vertex corrections are introduced for dynamics of
critical bosonic modes \cite{Garst_Namatic_QCP}. Since the fermion
self-energy is driven by $z = 3$ critical fluctuations, the
self-energy correction in the transverse mode, corresponding to
the $z = 2$ dynamics in the RPA level, gives rise to an additional
singular correction, changing the dynamical critical exponent from
$z = 2$ to $ z = 12/5$. This weird result is cured by the
Maki-Thompson vertex correction, where the singular correction of
the self-energy is cancelled by the singular vertex correction
exactly, which is a fundamental cancellation based on the Ward
identity.

The self-consistency of the Nambu-Eliashberg theory is far from
triviality for multi-scale quantum criticality, and this effective
field theory is certainly beyond the simple Eliashberg description
in respect that vertex corrections are introduced in some sense.
However, it turns out that critical exponents for thermodynamics
is completely the same as the conventional Eliashberg theory
without an anomalous self-energy correction \cite{Kim_AL}.



\subsection{Stability of the Eliashberg framework}

The Eliashberg theory is non-perturbative in respect that quantum
corrections are introduced fully self-consistently in the one-loop
level, consistent with one-loop renormalization group analysis.
Indeed, the scaling expression of the free energy was derived
explicitly, based on the Eliashberg approximation \cite{Kim_LW}.
In this respect the Eliashberg theory for quantum criticality
implies the theory with critical exponents given by the Eliashberg
approximation, satisfying the scaling theory.

One cautious person may criticize the Eliashberg approximation,
neglecting vertex corrections for self-energies. Recently, it has
been clarified that two dimensional Fermi surface problems are
still strongly interacting even in the large-$N$ limit
\cite{SungSik_Genus,Metlitski_Sachdev1,Metlitski_Sachdev2,McGreevy,Kim_Ladder,Kim_AL},
implying that vertex corrections should be incorporated. An
important question is whether these vertex corrections give rise
to novel critical exponents beyond the Eliashberg theory. Several
perturbative analysis demonstrated that although ladder-type
vertex corrections do not change critical exponents of the
Eliashberg theory, Aslamasov-Larkin corrections result in
modifications for such critical exponents
\cite{Metlitski_Sachdev1,Metlitski_Sachdev2}. If this is a general
feature beyond this level of approximation, various quantum
critical phenomena \cite{HFQCP} should be reconsidered because
novel anomalous exponents in fermion self-energy corrections are
expected to cause various novel critical exponents for
thermodynamics, transport, and etc. However, considering our
recent experiences on comparison with various experiments in heavy
fermion quantum criticality, we are surprised at the fact that
critical exponents based on the Eliashberg theory explain
thermodynamics \cite{Kim_Adel_Pepin}, both electrical and thermal
transport coefficients \cite{Kim_TR}, uniform spin susceptibility
\cite{Kim_Jia}, and Seebeck effect \cite{Kim_Pepin_Seebeck} quite
well.

Recently, one of the authors investigated the role of vertex
corrections non-perturbatively, summing vertex corrections up to
an infinite order \cite{Kim_Ladder,Kim_AL}. It turns out that
particular vertex corrections given by ladder diagrams do not
change Eliashberg critical exponents at all \cite{Kim_Ladder},
consistent with the perturbative analysis. This was performed in a
fully self-consistent way, resorting to the Ward identity. In
contrast with the previous perturbative analysis, Aslamasov-Larkin
corrections were shown not to modify the Eliashberg dynamics
\cite{Kim_AL}. Resorting to the analogy with superconductivity,
where the superconducting instability described by the
Aslamasov-Larkin vertex correction is reformulated by the
anomalous self-energy in the Eliashberg framework of the Nambu
spinor representation \cite{BCS_Book}, we claimed that the
off-diagonal self-energy associated with the 2$k_{F}$
particle-hole channel incorporates the same (Aslamasov-Larkin)
class of quantum corrections in the Nambu spinor representation.
We evaluated an anomalous pairing self-energy in the
Nambu-Eliashberg approximation, which vanishes at zero energy but
displays the same power law dependence for frequency as the normal
Eliashberg self-energy. As a result, even the pairing self-energy
correction does not modify the Eliashberg dynamics without the
Nambu spinor representation, resulting in essentially the same
scaling physics for thermodynamics at quantum criticality.
However, we admit that this issue should be investigated more
sincerely, resorting to a direct summation for such a class of
quantum corrections.

\section{Artificial electric field in the energy-momentum space}

An essential question is on the role of the anomalous self-energy
in transport. Considering that the off-diagonal self-energy has
the same momentum dependence as the kinetic-energy term, it is
natural to examine its role in the Hall conductivity.

An effective equation of motion for quasi-particle dynamics was
proposed as follows \cite{Shindou_E} \bqa && \frac{d
\boldsymbol{R}}{d t} = \boldsymbol{v} + (\boldsymbol{\mathcal{B}}
- \boldsymbol{\mathcal{E}} \times \boldsymbol{v}) \times \frac{d
\boldsymbol{k}}{d t} , \nn && \frac{d \boldsymbol{k}}{d t} = -
\boldsymbol{E} + \boldsymbol{B} \times \frac{d \boldsymbol{R}}{d
t} , \eqa describing the wave-packet dynamics in the
semi-classical level. $\boldsymbol{R}$ represents the center
coordinate of the wave-packet and $\boldsymbol{k}$ expresses the
Bloch wave vector. $\boldsymbol{E}$ and $\boldsymbol{B}$ are
external electric and magnetic fields while
$\boldsymbol{\mathcal{E}}$ and $\boldsymbol{\mathcal{B}}$ are
artificial electric and magnetic fields in the energy and momentum
space. $\boldsymbol{\mathcal{B}}$ is usually referred as the Berry
curvature. $\boldsymbol{v} = \frac{\partial
E_{\boldsymbol{k}}}{\partial \boldsymbol{k}}$ is the group
velocity, where $E_{\boldsymbol{k}}$ is the quasi-particle energy
dispersion. A key ingredient is that interactions give rise to an
artificial electric field in the energy-momentum space, modifying
the quasi-particle dynamics according to the Maxwell equation. As
a result, the Hall conductivity should be corrected in the
following way \cite{Shindou_E} \bqa && \sigma_{xy} =
\frac{e^{2}}{\hbar} \int \frac{d^{2} \boldsymbol{k}}{(2\pi)^{2}}
f(E_{\boldsymbol{k}}) \Bigl( \mathcal{B}_{z} - [ v_{y}
\mathcal{E}_{x} - v_{x} \mathcal{E}_{y} ] \Bigr) , \eqa where the
Berry curvature is shifted by $\boldsymbol{\mathcal{E}}\times
\boldsymbol{v}$ due to interactions. $f(E_{\boldsymbol{k}})$ is
the Fermi-Dirac distribution function. In this section we
calculate this quantity with an introduction of the anomalous
self-energy correction.

We consider the following effective Hamiltonian $H_{eff} =
\boldsymbol{N}(\boldsymbol{k},\omega) \cdot \boldsymbol{\sigma}$,
which incorporates the effect of interactions through the
self-energy. The pseudospin vector is given by \bqa &&
\boldsymbol{N}(\boldsymbol{k},\omega) =
\boldsymbol{d}(\boldsymbol{k}) + \Delta \boldsymbol{d} (\omega) ,
\nn && \boldsymbol{d}(\boldsymbol{k}) = - v_{f} k_{y}
\boldsymbol{\hat{\sigma}}_{x} + v_{f} k_{x}
\boldsymbol{\hat{\sigma}}_{y} + d_{z} (\boldsymbol{k})
\boldsymbol{\hat{\sigma}}_{z} , \nn && \Delta
\boldsymbol{d}(\omega) = - \Phi(\omega) (k_{y}/k)
\boldsymbol{\hat{\sigma}}_{x} + \Phi(\omega) (k_{x}/k)
\boldsymbol{\hat{\sigma}}_{y} + \Delta d_{z}(\omega)
\boldsymbol{\hat{\sigma}}_{z} , \nn \eqa where the
$\boldsymbol{d}(\boldsymbol{k})$ vector is nothing but the bare
dispersion and $\Delta \boldsymbol{d}(\omega)$ is associated with
the anomalous self-energy correction. $d_{z} (\boldsymbol{k})$ is
introduced for time reversal symmetry breaking as the
$z$-directional magnetic field, set to be $d_{z} (\boldsymbol{k})
\rightarrow m$. $\Delta d_{z}(\omega)$ is also introduced for
consistency. $k = \sqrt{k_{x}^{2} + k_{y}^{2}}$ is the amplitude
of the momentum.

The Berry curvature and artificial electric field are expressed in
terms of the pseudospin vector \bqa && \mathcal{B}_{z} =
\frac{1}{2}
\partial_{k_x} \boldsymbol{\hat{N}} \times
\partial_{k_y} \boldsymbol{\hat{N}} \cdot \boldsymbol{\hat{N}} , \nn && \mathcal{E}_{x} =
\frac{1}{2} \partial_{k_y} \boldsymbol{\hat{N}} \times
\partial_{\omega} \boldsymbol{\hat{N}} \cdot \boldsymbol{\hat{N}} , \nn &&
\mathcal{E}_{y} = \frac{1}{2} \partial_{\omega}
\boldsymbol{\hat{N}} \times
\partial_{k_x} \boldsymbol{\hat{N}} \cdot \boldsymbol{\hat{N}} ,
\eqa where $\boldsymbol{\hat{N}} = \boldsymbol{N} /
|\boldsymbol{N}|$ is the unit vector.

One can understand this expression based on another equivalent
representation, where the field strength is expressed by the gauge
field. Resorting to the CP$^{1}$ representation \cite{Spin_Book}
\bqa && \boldsymbol{N} \cdot \boldsymbol{\sigma} =
|\boldsymbol{N}| \boldsymbol{U} \boldsymbol{\sigma}_{3}
\boldsymbol{U}^{\dagger} , \eqa where $\boldsymbol{U}$ is an SU(2)
matrix field to describe a direction of the pseudospin vector, we
can define an SU(2) gauge connection \bqa &&
\boldsymbol{\mathcal{A}}_{\mu} = - i [\partial_{\mu}
\boldsymbol{U}^{\dagger}] \boldsymbol{U} . \eqa This
naturally leads to both Berry curvature and artificial electric
field \bqa && \mathcal{B}_{z} = \partial_{k_x} \mathcal{A}_{y} -
\partial_{k_y} \mathcal{A}_{x} , \nn && \mathcal{E}_{x} =
\partial_{k_y} \mathcal{A}_{\omega} - \partial_{\omega}
\mathcal{A}_{k_y} , \nn && \mathcal{E}_{y} = \partial_{\omega}
\mathcal{A}_{k_x} -
\partial_{k_x} \mathcal{A}_{\omega} ,  \eqa where the U(1)
projected component of the SU(2) gauge field is given by
$\mathcal{A}_{\mu} = \boldsymbol{\mathcal{A}}_{\mu}^{(3)}$.

%
Inserting the pseudospin vector Eq. (12) into Eq. (13),
we obtain both Berry curvature and artificial electric field \bqa
&& \mathcal{B}_{z} \approx \frac{1}{2} \frac{[ v_{f} +
\Phi(\omega)/k ]^{2} [m + \Delta d_{z}(\omega)]}{\Bigl( [ v_{f} k
+ \Phi(\omega) ]^{2} + [m + \Delta d_{z}(\omega)]^{2}
\Bigr)^{3/2}} , \nn && \mathcal{E}_{x} \approx \frac{1}{2} k_{x}
\mathcal{F} , ~~~~~ \mathcal{E}_{y} \approx \frac{1}{2} k_{y}
\mathcal{F} , \nonumber \eqa where $\mathcal{F}$ is given by \bqa
&& \mathcal{F} = - \frac{[\partial_{\omega} \Phi(\omega)] (1/k) [
v_{f} + \Phi(\omega)/k ] [m + \Delta d_{z}(\omega)]}{\Bigl( [
v_{f} k + \Phi(\omega) ]^{2} + [m + \Delta d_{z}(\omega)]^{2}
\Bigr)^{3/2}} \nn && + \frac{ [ v_{f} + \Phi(\omega)/k ]^{2}
[\partial_{\omega} \Delta d_{z}(\omega)]}{\Bigl( [ v_{f} k +
\Phi(\omega) ]^{2} + [m + \Delta d_{z}(\omega)]^{2} \Bigr)^{3/2}}
. \nonumber \eqa

Considering the group velocity \bqa && v_{y} = k_{y} \mathcal{V} ,
~~~~~ v_{x} = k_{x} \mathcal{V} , \nonumber \eqa with \bqa &&
\mathcal{V} = \frac{1}{k} \frac{[ v_{f} k + \Phi(\omega) ] v_{f}
}{\sqrt{ [ v_{f} k + \Phi(\omega) ]^{2} + [m + \Delta
d_{z}(\omega)]^{2}}} , \nonumber \eqa we find that the
contribution of the electric field $\boldsymbol{\mathcal{E}}$
to the shift of $\mathcal{B}$ in Eq. (11)
vanishes identically \bqa && v_{y} \mathcal{E}_{x} -
v_{x} \mathcal{E}_{y} = 0 . \nonumber \eqa On the other hand, the
Berry curvature should be modified in the presence of the
self-energy correction. As a result, we obtain the following
expression for the Hall conductivity
\begin{widetext}
\bqa && \sigma_{xy} = \frac{1}{2} \frac{e^{2}}{h}
\int_{0}^{\infty} d k \frac{1}{k} \frac{[ v_{f} k + \Phi(\omega)
]^{2} [m + \Delta d_{z}(\omega)]}{\Bigl( [ v_{f} k + \Phi(\omega)
]^{2} + [m + \Delta d_{z}(\omega)]^{2} \Bigr)^{3/2}} = \frac{1}{2}
\frac{e^{2}}{h} \Bigl( F[k_{H} \rightarrow \infty] - F[k_{L}
\rightarrow 0] \Bigr) , \eqa where \bqa && F[k_{H} \rightarrow
\infty] = \frac{ [m + \Delta d_{z}(\omega)] \Phi(\omega)
}{\Bigl([m + \Delta d_{z}(\omega)]^{2} + \Phi^{2}(\omega) \Bigr) }
+ \frac{ [m + \Delta d_{z}(\omega)] \Phi^{2}(\omega) }{\Bigl(
\Phi^{2}(\omega) + [m + \Delta d_{z}(\omega)]^{2} \Bigr)^{3/2}}
\ln \Bigl( \frac{1}{ \Phi(\omega) + \sqrt{\Phi^{2}(\omega) + [m +
\Delta d_{z}(\omega)]^{2}}  } \Bigr) , \nn && F[k_{L} \rightarrow
0] = \frac{- [m + \Delta d_{z}(\omega)]^{3} + [m + \Delta
d_{z}(\omega)] \Phi^{2}(\omega) }{\Bigl( \Phi^{2}(\omega) + [m +
\Delta d_{z}(\omega)]^{2} \Bigr)^{3/2}} \nn && + \frac{ [m +
\Delta d_{z}(\omega)] \Phi^{2}(\omega) }{\Bigl( \Phi^{2}(\omega) +
[m + \Delta d_{z}(\omega)]^{2} \Bigr)^{3/2}} \ln \Bigl(
\frac{k_{L}}{[m + \Delta d_{z}(\omega)]^{2} + \Phi^{2}(\omega) +
\sqrt{\Phi^{2}(\omega) + [m + \Delta d_{z}(\omega)]^{2}} \sqrt{
\Phi^{2}(\omega) + [m + \Delta d_{z}(\omega)]^{2}} } \Bigr) .
\nonumber \eqa
\end{widetext}
It is straightforward to see that the noninteracting case in
$\Phi(\omega) = 0$ recovers the well known result, that is, the
half Hall conductivity \cite{Review_TI}, where $F[k_{H}
\rightarrow \infty] = 0$ and $F[k_{L} \rightarrow 0] = -
\boldsymbol{sgn}(m)$.

An interesting point in this expression is that the Hall
conductivity diverges in the $k_{L} \rightarrow 0$ limit, seen
from the log term. Although we do not understand the reason for
this divergence, it implies that we should incorporate vertex
corrections even for the intrinsic topological effect in the Hall
conductivity when interactions are introduced. If we take into
account vertex corrections, this log divergence may be re-summed
to result in the common power-law behavior with an exponent, which
vanishes in the $k_{L} \rightarrow 0$ limit.

We suspect that the underlying mechanism for this divergence may
be charged vortices. As discussed before, the parity anomaly
assigns an electric charge to a vortex excitation
\cite{Ryu_Dirac_Mass}. From the ordered side in the duality picture,
such vortex excitations are well defined particles, thus
taken into account with electrons on an equal footing. An
interplay between electrons and vortices would be described by the
mutual Chern-Simons theory \cite{Review_TI}, where their mutual
statistics is guaranteed by the mutual Chern-Simons term. It is
not clear at all how such an interplay affects transport, in
particular, the Hall conductivity. It will be an interesting
future direction to investigate this problem in the duality
picture because topological excitations carry nontrivial fermion
quantum numbers in this situation.

\section{Discussion and Summary}

In this study we constructed an Eliashberg framework for the
Hertz-Moriya-Millis theory \cite{Rech_FMQCP}, describing
ferromagnetic quantum criticality in the surface state of three
dimensional topological insulators. An idea is to introduce an
anomalous self-energy correction beyond the previous study
\cite{Xu_RPA}, resulting from spin-flip scattering. We could show
that our ansatz for the anomalous self-energy allows a set of
fully self-consistent solutions for electrons and spin
fluctuations.

The spin-fluctuation dynamics was shown to coincide with the
solution based on the random phase approximation, where both $z =
3$ and $z = 2$ quantum critical dynamics arise \cite{Xu_RPA}. The
$z = 3$ quantum criticality describes the longitudinal spin
dynamics, where only forward scattering contributions result in
the $z = 3$ Landau damping, while the $z = 2$ quantum criticality
for transverse spin fluctuations originates from the spin-flip
scattering, involved with the finite-momentum transfer due to the
spin-orbital quenching.

This multi-scale quantum criticality was shown to allow the
self-consistent Nambu-Eliashberg solution for the electron
dynamics. The normal self-energy turns out to follow the $z = 3$
quantum criticality because such critical fluctuations give rise
to more singular contributions than the $z = 2$ transverse mode.
We found that the anomalous self-energy shows exactly the same
frequency dependence as the normal self-energy although the
underlying mechanism is not completely clear. The equivalence
between the normal and anomalous self-energies with a different
sign is at the heart of the self-consistency in the
Nambu-Eliashberg theory for multi-scale quantum criticality. In
other words, the simple Eliashberg theory without the off diagonal
self-energy cannot be self-consistent for multi-scale quantum
criticality. We argued that the mechanism of cancellation between
the normal and anomalous self-energies in the fermion bubble
diagram may be rooted in the Ward identity, comparing our result
with the perturbative analysis for nematic quantum criticality,
where the singular self-energy correction is cancelled by the
vertex correction based on the Ward identity
\cite{Garst_Namatic_QCP}.

We demonstrated that such an off-diagonal self-energy gives rise
to an artificial electric field in the energy-momentum space
beyond the static Berry curvature from the topological band
structure. In particular, we calculated the anomalous Hall
conductivity in terms of both Berry curvature and artificial
electric field. Although the contribution from the artificial
electric field vanishes identically, we could find that the Berry
curvature is modified due to the anomalous self-energy correction,
resulting in the log-divergence for the Hall conductivity. We
argued that the log-divergence may disappear if vertex corrections
are re-summed up to an infinite order, which is expected to turn
the log-divergence into the power-law dependence, vanishing
algebraically. We suggested an idea that this phenomenology may be
related with charged vortices in the duality picture, where the
electron quantum number of the vortex results from the quantum
anomaly, an essential feature of the topological surface state.

Before closing, we discuss the stability of the Eliashberg
approximation in another aspect, considering an interesting recent
study for nematic quantum criticality \cite{Garst_Nematic}. The
nematic quantum critical point is well known to show multi-scale
quantum criticality, where critical dynamics of nematic
fluctuations is governed by both $z = 3$ and $z = 2$.
%
%
An important notice is that although the self-energy correction
results from scattering with $z = 3$ critical fluctuations
dominantly, scattering with $z = 2$ critical modes gives rise to
the logarithmic singularity for the quasi-particle residue
\cite{Garst_Nematic}, which means that vertex corrections should
be introduced up to an infinite order. Based on a complicated
renormalization group argument, the authors proposed an
interesting expression for the electron Green's function \bqa &&
G(\boldsymbol{k},i\omega) \propto \frac{ D_{0}^{-\eta}}{[i\omega +
\Sigma(i\omega) - \epsilon_{\boldsymbol{k}}]^{1-\eta}} . \nonumber
\eqa $D_{0}$ has a constant of an energy unit proportional to the
Fermi energy, and $\Sigma(i\omega) \propto |\omega|^{2/3}$ is the
$z = 3$ self-energy. $\epsilon_{\boldsymbol{k}}$ is the band
dispersion. It should be noted that the anomalous exponent $\eta$
results from scattering with $z = 2$ critical fluctuations.

In the present system a significant quantity is the off diagonal
self-energy. Although such an anomalous self-energy will not be
allowed in nematic quantum criticality due to the absence of the
spin-orbit coupling, it may play an important role for the
exponent $\eta$ in ferromagnetic quantum criticality of the
topological surface. A more complete framework would be to
introduce vertex corrections fully self-consistently with
self-energy corrections in the Nambu-spinor representation, where
the off-diagonal self-energy contribution is also included.

In summary, quantum criticality with the topological band
structure is expected to open a novel direction of research in
respect that the interplay between interaction and topology may
cause unexpected non-Fermi liquid physics. We found an anomalous
behavior in the intrinsic topological contribution of the Hall
coefficient, the source of which is the energy-momentum-space
field strength beyond the static Berry curvature as a result of
the interplay between quantum criticality and topological
structure.

\begin{acknowledgments}

We would like to thank S. Ryu for his eight-hours lecture on
topological insulators in APCTP. K.-S. Kim was supported by the
National Research Foundation of Korea (NRF) grant funded by the
Korea government (MEST) (No. 2010-0074542).

\end{acknowledgments}

\appendix

\begin{widetext}

\section{Polarization function}

In appendix A we derive the self-energy correction for spin
fluctuations, where the longitudinal spin dynamics is given by $z
= 3$ while the transverse one is described by $z = 2$. Inserting
the electron Green's function with the ansatz for the electron
self-energy matrix into the Eliashberg equation for the boson
self-energy, we obtain \bqa &&
\boldsymbol{\Pi}_{ij}(\boldsymbol{q},i\Omega) = \frac{g^{2}}{2}
\frac{1}{\beta} \sum_{i\omega} \int \frac{d^{2}
\boldsymbol{k}}{(2\pi)^{2}} \boldsymbol{tr} \Bigl\{
\boldsymbol{\sigma}_{i} \frac{[i\omega+i\Omega + \mu -
\Sigma(i\omega+i\Omega)] \boldsymbol{I} + [v_{f} |\boldsymbol{k} +
\boldsymbol{q}| + \Phi(i\omega+i\Omega) ] \epsilon_{nm}
(\boldsymbol{\hat{k}}_{n} + \boldsymbol{\hat{q}}_{n} )
\boldsymbol{\sigma}_{m}}{[ i\omega+i\Omega + \mu -
\Sigma(i\omega+i\Omega)]^{2} - [v_{f} |\boldsymbol{k} +
\boldsymbol{q}| + \Phi(i\omega+i\Omega) ]^{2} } \nn &&
\boldsymbol{\sigma}_{j} \frac{[i \omega + \mu - \Sigma(i\omega)]
\boldsymbol{I} + [v_{f} |\boldsymbol{k}| + \Phi(i\omega)
]\epsilon_{n'm'} \boldsymbol{\hat{k}}_{n'}
\boldsymbol{\sigma}_{m'}}{[i \omega + \mu - \Sigma(i\omega)]^{2} -
[v_{f} |\boldsymbol{k}| + \Phi(i\omega) ]^{2} } \Bigr\}   , \eqa
where the following approximation for the momentum \bqa &&
\boldsymbol{\hat{k}}_{n} + \boldsymbol{\hat{q}}_{n} \equiv
\frac{\boldsymbol{k}_{n} + \boldsymbol{q}_{n}}{|\boldsymbol{k} +
\boldsymbol{q} |} \approx
\frac{\boldsymbol{k}_{n}}{|\boldsymbol{k}_{F}|} +
\frac{\boldsymbol{q}_{n}}{|\boldsymbol{k}_{F}|} \eqa is utilized.

Resorting to identities for Pauli matrix products \bqa &&
\boldsymbol{\sigma}_{i} \boldsymbol{\sigma}_{j} = \delta_{ij}
\boldsymbol{I} + i \epsilon_{ijl} \boldsymbol{\sigma}_{l} , \nn &&
\boldsymbol{\sigma}_{i} \boldsymbol{\sigma}_{m}
\boldsymbol{\sigma}_{j} = \delta_{im} \boldsymbol{\sigma}_{j} + i
\epsilon_{imj} \boldsymbol{I} - (\delta_{ij} \delta_{ms} -
\delta_{is} \delta_{mj}) \boldsymbol{\sigma}_{s} , \nn &&
\boldsymbol{\sigma}_{i} \boldsymbol{\sigma}_{m}
\boldsymbol{\sigma}_{j} \boldsymbol{\sigma}_{m'} = \delta_{im}
\Bigl( \delta_{jm'} \boldsymbol{I} + i \epsilon_{jm'l'}
\boldsymbol{\sigma}_{l'} \Bigr) + i \epsilon_{imj}
\boldsymbol{\sigma}_{m'} - (\delta_{ij} \delta_{ms} - \delta_{is}
\delta_{mj}) \Bigl( \delta_{sm'} \boldsymbol{I} + i
\epsilon_{sm'l'} \boldsymbol{\sigma}_{l'} \Bigr)   \eqa and
performing the decomposition for the denominator, we can rewrite
Eq. (A1) as follows \bqa &&
\boldsymbol{\Pi}_{ij}(\boldsymbol{q},i\Omega) = N_{\sigma} g^{2}
\frac{1}{\beta} \sum_{i\omega} \int \frac{d^{2}
\boldsymbol{k}}{(2\pi)^{2}} \frac{1}{2[i\omega+i\Omega + \mu -
\Sigma(i\omega+i\Omega)]} \frac{1}{2[i\omega + \mu -
\Sigma(i\omega)] } \nn && \Bigl\{ \mathcal{G}[i \omega + i \Omega,
v_{f} |\boldsymbol{k} + \boldsymbol{q}| + \Phi(i\omega+i\Omega)]
\mathcal{G}[i \omega, v_{f} |\boldsymbol{k}| + \Phi(i\omega)] +
\mathcal{G}[i \omega + i \Omega, - v_{f} |\boldsymbol{k} +
\boldsymbol{q}| - \Phi(i\omega+i\Omega)] \mathcal{G}[i \omega,
v_{f} |\boldsymbol{k}| + \Phi(i\omega)] \nn && + \mathcal{G}[i
\omega + i \Omega, v_{f} |\boldsymbol{k} + \boldsymbol{q}| +
\Phi(i\omega+i\Omega)] \mathcal{G}[i \omega, - v_{f}
|\boldsymbol{k}| - \Phi(i\omega)] + \mathcal{G}[i \omega + i
\Omega, - v_{f} |\boldsymbol{k} + \boldsymbol{q}| -
\Phi(i\omega+i\Omega)] \mathcal{G}[i \omega, - v_{f}
|\boldsymbol{k}| - \Phi(i\omega)] \Bigr\} \nn && \Bigl\{ \Bigl(
[i\omega+i\Omega + \mu - \Sigma(i\omega+i\Omega)] [i \omega + \mu
- \Sigma(i\omega)] - [v_{f} |\boldsymbol{k} + \boldsymbol{q}| +
\Phi(i\omega+i\Omega) ] [v_{f} |\boldsymbol{k}| + \Phi(i\omega) ]
(\boldsymbol{\hat{k}} + \boldsymbol{\hat{q}} ) \cdot
\boldsymbol{\hat{k}} \Bigr) \delta_{ij} \nn && + [v_{f}
|\boldsymbol{k} + \boldsymbol{q}| + \Phi(i\omega+i\Omega) ] [v_{f}
|\boldsymbol{k}| + \Phi(i\omega) ] [\epsilon_{in} \epsilon_{jn'}
(\boldsymbol{\hat{k}}_{n} + \boldsymbol{\hat{q}}_{n} )
\boldsymbol{\hat{k}}_{n'} + \epsilon_{jn} \epsilon_{in'}
(\boldsymbol{\hat{k}}_{n} + \boldsymbol{\hat{q}}_{n} )
\boldsymbol{\hat{k}}_{n'}] \Bigr\} , \eqa where $\epsilon_{ij}$ is
an antisymmetric tensor with $i,j = x,y$, and \bqa &&
\mathcal{G}[i \omega, \mathcal{X}] = \frac{1}{ i\omega+ \mu -
\Sigma(i\omega) - \mathcal{X} } . \nonumber \eqa

We rearrange the above expression as \bqa &&
\boldsymbol{\Pi}_{ij}(\boldsymbol{q},i\Omega) = N_{\sigma}
\frac{g^{2}}{4} \frac{1}{\beta} \sum_{i\omega} \int \frac{d^{2}
\boldsymbol{k}}{(2\pi)^{2}} \delta_{ij} \nn && \Bigl\{
\mathcal{G}[i \omega + i \Omega, v_{f} |\boldsymbol{k} +
\boldsymbol{q}| + \Phi(i\omega+i\Omega)] \mathcal{G}[i \omega,
v_{f} |\boldsymbol{k}| + \Phi(i\omega)] + \mathcal{G}[i \omega + i
\Omega, - v_{f} |\boldsymbol{k} + \boldsymbol{q}| -
\Phi(i\omega+i\Omega)] \mathcal{G}[i \omega, v_{f}
|\boldsymbol{k}| + \Phi(i\omega)] \nn && + \mathcal{G}[i \omega +
i \Omega, v_{f} |\boldsymbol{k} + \boldsymbol{q}| +
\Phi(i\omega+i\Omega)] \mathcal{G}[i \omega, - v_{f}
|\boldsymbol{k}| - \Phi(i\omega)] + \mathcal{G}[i \omega + i
\Omega, - v_{f} |\boldsymbol{k} + \boldsymbol{q}| -
\Phi(i\omega+i\Omega)] \mathcal{G}[i \omega, - v_{f}
|\boldsymbol{k}| - \Phi(i\omega)] \Bigr\} \nn && + N_{\sigma}
\frac{g^{2}}{4} \frac{1}{\beta} \sum_{i\omega} \int \frac{d^{2}
\boldsymbol{k}}{(2\pi)^{2}} \Bigl\{ - (\boldsymbol{\hat{k}} +
\boldsymbol{\hat{q}} ) \cdot \boldsymbol{\hat{k}} \delta_{ij} +
[\epsilon_{in} \epsilon_{jn'} (\boldsymbol{\hat{k}}_{n} +
\boldsymbol{\hat{q}}_{n} ) \boldsymbol{\hat{k}}_{n'} +
\epsilon_{jn} \epsilon_{in'} (\boldsymbol{\hat{k}}_{n} +
\boldsymbol{\hat{q}}_{n} ) \boldsymbol{\hat{k}}_{n'}] \Bigr\} \nn
&& \Bigl\{ \mathcal{G}[i \omega + i \Omega, v_{f} |\boldsymbol{k}
+ \boldsymbol{q}| + \Phi(i\omega+i\Omega)] \mathcal{G}[i \omega,
v_{f} |\boldsymbol{k}| + \Phi(i\omega)] - \mathcal{G}[i \omega + i
\Omega, - v_{f} |\boldsymbol{k} + \boldsymbol{q}| -
\Phi(i\omega+i\Omega)] \mathcal{G}[i \omega, v_{f}
|\boldsymbol{k}| + \Phi(i\omega)] \nn && - \mathcal{G}[i \omega +
i \Omega, v_{f} |\boldsymbol{k} + \boldsymbol{q}| +
\Phi(i\omega+i\Omega)] \mathcal{G}[i \omega, - v_{f}
|\boldsymbol{k}| - \Phi(i\omega)] + \mathcal{G}[i \omega + i
\Omega, - v_{f} |\boldsymbol{k} + \boldsymbol{q}| -
\Phi(i\omega+i\Omega)] \mathcal{G}[i \omega, - v_{f}
|\boldsymbol{k}| - \Phi(i\omega)] \Bigr\}  , \nn \eqa where the
frequency term in the numerator is reorganized.

It is convenient for the momentum integration to linearize the
band dispersion near the chemical potential \bqa && \mu - v_{f}
|\boldsymbol{k} + \boldsymbol{q}| \approx - v_{f}
\boldsymbol{\hat{k}}_{f} \cdot (\boldsymbol{k} -
\boldsymbol{k}_{f}) -  v_{f} \boldsymbol{\hat{k}}_{f} \cdot
\boldsymbol{q}  ,  \eqa where the Fermi momentum
$\boldsymbol{k}_{f} = |\boldsymbol{k}_{f}|
\boldsymbol{\hat{k}}_{f}$ is defined from $\mu - v_{f}
|\boldsymbol{k}_{f}| = 0$.

Inserting the linearized band into Eq. (A5) and performing the
decomposition for the denominator, we obtain \bqa &&
\boldsymbol{\Pi}_{ij}(\boldsymbol{q},i\Omega) \approx N_{\sigma}
\frac{g^{2}}{4} \frac{1}{\beta} \sum_{i\omega} \int \frac{d^{2}
\boldsymbol{k}}{(2\pi)^{2}} \delta_{ij} \frac{1}{i\Omega -
[\Sigma(i\omega+i\Omega) - \Sigma(i\omega)] -
[\Phi(i\omega+i\Omega) - \Phi(i\omega)] -  v_{f}
\boldsymbol{\hat{k}}_{f} \cdot \boldsymbol{q}} \nn && \Bigl\{
\frac{1}{ i\omega - \Sigma(i\omega) - \Phi(i\omega) - v_{f}
\boldsymbol{\hat{k}}_{f} \cdot (\boldsymbol{k} -
\boldsymbol{k}_{f})} - \frac{1}{ i\omega+i\Omega -
\Sigma(i\omega+i\Omega) - \Phi(i\omega+i\Omega) - v_{f}
\boldsymbol{\hat{k}}_{f} \cdot (\boldsymbol{k} -
\boldsymbol{k}_{f}) -  v_{f} \boldsymbol{\hat{k}}_{f} \cdot
\boldsymbol{q}} \Bigr\} \nn && + N_{\sigma} \frac{g^{2}}{4}
\frac{1}{\beta} \sum_{i\omega} \int \frac{d^{2}
\boldsymbol{k}}{(2\pi)^{2}} \Bigl\{- (\boldsymbol{\hat{k}} +
\boldsymbol{\hat{q}} ) \cdot \boldsymbol{\hat{k}} \delta_{ij} +
[\epsilon_{in} \epsilon_{jn'} (\boldsymbol{\hat{k}}_{n} +
\boldsymbol{\hat{q}}_{n} ) \boldsymbol{\hat{k}}_{n'} +
\epsilon_{jn} \epsilon_{in'} (\boldsymbol{\hat{k}}_{n} +
\boldsymbol{\hat{q}}_{n} ) \boldsymbol{\hat{k}}_{n'}] \Bigr\} \nn
&& \frac{1}{i\Omega - [\Sigma(i\omega+i\Omega) - \Sigma(i\omega)]
- [\Phi(i\omega+i\Omega) - \Phi(i\omega)] -  v_{f}
\boldsymbol{\hat{k}}_{f} \cdot \boldsymbol{q}} \nn && \Bigl\{
\frac{1}{ i\omega - \Sigma(i\omega) - \Phi(i\omega) - v_{f}
\boldsymbol{\hat{k}}_{f} \cdot (\boldsymbol{k} -
\boldsymbol{k}_{f})} - \frac{1}{ i\omega+i\Omega -
\Sigma(i\omega+i\Omega) - \Phi(i\omega+i\Omega) - v_{f}
\boldsymbol{\hat{k}}_{f} \cdot (\boldsymbol{k} -
\boldsymbol{k}_{f}) -  v_{f} \boldsymbol{\hat{k}}_{f} \cdot
\boldsymbol{q}} \Bigr\} , \eqa where only dominant contributions
are selected, which means that all terms with the chemical
potential in the denominator are safely neglected.

Before evaluating integrals, we summarize each component for the
spin-fluctuation self-energy as \bqa &&
\Pi_{xx}(\boldsymbol{q},i\Omega) = N_{\sigma} \frac{g^{2}}{4}
\frac{1}{\beta} \sum_{i\omega} \int \frac{d^{2}
\boldsymbol{k}}{(2\pi)^{2}} \frac{1}{i\Omega -
[\Sigma(i\omega+i\Omega) - \Sigma(i\omega)] -
[\Phi(i\omega+i\Omega) - \Phi(i\omega)] -  v_{f}
\boldsymbol{\hat{k}}_{f} \cdot \boldsymbol{q}} \nn && \Bigl\{
\frac{1}{ i\omega - \Sigma(i\omega) - \Phi(i\omega) - v_{f}
\boldsymbol{\hat{k}}_{f} \cdot (\boldsymbol{k} -
\boldsymbol{k}_{f})} - \frac{1}{ i\omega+i\Omega -
\Sigma(i\omega+i\Omega) - \Phi(i\omega+i\Omega) - v_{f}
\boldsymbol{\hat{k}}_{f} \cdot (\boldsymbol{k} -
\boldsymbol{k}_{f}) -  v_{f} \boldsymbol{\hat{k}}_{f} \cdot
\boldsymbol{q}} \Bigr\} \nn && + N_{\sigma} \frac{g^{2}}{4}
\frac{1}{\beta} \sum_{i\omega} \int \frac{d^{2}
\boldsymbol{k}}{(2\pi)^{2}} \frac{- (\boldsymbol{\hat{k}} +
\boldsymbol{\hat{q}} ) \cdot \boldsymbol{\hat{k}} + 2
(\boldsymbol{\hat{k}}_{y} + \boldsymbol{\hat{q}}_{y} )
\boldsymbol{\hat{k}}_{y}}{i\Omega - [\Sigma(i\omega+i\Omega) -
\Sigma(i\omega)] - [\Phi(i\omega+i\Omega) - \Phi(i\omega)] - v_{f}
\boldsymbol{\hat{k}}_{f} \cdot \boldsymbol{q}} \nn && \Bigl\{
\frac{1}{ i\omega - \Sigma(i\omega) - \Phi(i\omega) - v_{f}
\boldsymbol{\hat{k}}_{f} \cdot (\boldsymbol{k} -
\boldsymbol{k}_{f})} - \frac{1}{ i\omega+i\Omega -
\Sigma(i\omega+i\Omega) - \Phi(i\omega+i\Omega) - v_{f}
\boldsymbol{\hat{k}}_{f} \cdot (\boldsymbol{k} -
\boldsymbol{k}_{f}) -  v_{f} \boldsymbol{\hat{k}}_{f} \cdot
\boldsymbol{q}} \Bigr\}  , \nn && \Pi_{yy}(\boldsymbol{q},i\Omega)
= N_{\sigma} \frac{g^{2}}{4} \frac{1}{\beta} \sum_{i\omega} \int
\frac{d^{2} \boldsymbol{k}}{(2\pi)^{2}} \frac{1}{i\Omega -
[\Sigma(i\omega+i\Omega) - \Sigma(i\omega)] -
[\Phi(i\omega+i\Omega) - \Phi(i\omega)] -  v_{f}
\boldsymbol{\hat{k}}_{f} \cdot \boldsymbol{q}} \nn && \Bigl\{
\frac{1}{ i\omega - \Sigma(i\omega) - \Phi(i\omega) - v_{f}
\boldsymbol{\hat{k}}_{f} \cdot (\boldsymbol{k} -
\boldsymbol{k}_{f})} - \frac{1}{ i\omega+i\Omega -
\Sigma(i\omega+i\Omega) - \Phi(i\omega+i\Omega) - v_{f}
\boldsymbol{\hat{k}}_{f} \cdot (\boldsymbol{k} -
\boldsymbol{k}_{f}) -  v_{f} \boldsymbol{\hat{k}}_{f} \cdot
\boldsymbol{q}} \Bigr\} \nn && + N_{\sigma} \frac{g^{2}}{4}
\frac{1}{\beta} \sum_{i\omega} \int \frac{d^{2}
\boldsymbol{k}}{(2\pi)^{2}} \frac{- (\boldsymbol{\hat{k}} +
\boldsymbol{\hat{q}} ) \cdot \boldsymbol{\hat{k}} + 2
(\boldsymbol{\hat{k}}_{x} + \boldsymbol{\hat{q}}_{x} )
\boldsymbol{\hat{k}}_{x}}{i\Omega - [\Sigma(i\omega+i\Omega) -
\Sigma(i\omega)] - [\Phi(i\omega+i\Omega) - \Phi(i\omega)] - v_{f}
\boldsymbol{\hat{k}}_{f} \cdot \boldsymbol{q}} \nn && \Bigl\{
\frac{1}{ i\omega - \Sigma(i\omega) - \Phi(i\omega) - v_{f}
\boldsymbol{\hat{k}}_{f} \cdot (\boldsymbol{k} -
\boldsymbol{k}_{f})} - \frac{1}{ i\omega+i\Omega -
\Sigma(i\omega+i\Omega) - \Phi(i\omega+i\Omega) - v_{f}
\boldsymbol{\hat{k}}_{f} \cdot (\boldsymbol{k} -
\boldsymbol{k}_{f}) -  v_{f} \boldsymbol{\hat{k}}_{f} \cdot
\boldsymbol{q}} \Bigr\} , \nn && \Pi_{xy}(\boldsymbol{q},i\Omega)
= \Pi_{yx}(\boldsymbol{q},i\Omega) = - N_{\sigma} \frac{g^{2}}{4}
\frac{1}{\beta} \sum_{i\omega} \int \frac{d^{2}
\boldsymbol{k}}{(2\pi)^{2}} \frac{2
\boldsymbol{\hat{k}}_{x}\boldsymbol{\hat{k}}_{y} +
(\boldsymbol{\hat{q}}_{x} \boldsymbol{\hat{k}}_{y} +
\boldsymbol{\hat{q}}_{y} \boldsymbol{\hat{k}}_{x} ) }{i\Omega -
[\Sigma(i\omega+i\Omega) - \Sigma(i\omega)] -
[\Phi(i\omega+i\Omega) - \Phi(i\omega)] - v_{f}
\boldsymbol{\hat{k}}_{f} \cdot \boldsymbol{q}} \nn && \Bigl\{
\frac{1}{ i\omega - \Sigma(i\omega) - \Phi(i\omega) - v_{f}
\boldsymbol{\hat{k}}_{f} \cdot (\boldsymbol{k} -
\boldsymbol{k}_{f})} - \frac{1}{ i\omega+i\Omega -
\Sigma(i\omega+i\Omega) - \Phi(i\omega+i\Omega) - v_{f}
\boldsymbol{\hat{k}}_{f} \cdot (\boldsymbol{k} -
\boldsymbol{k}_{f}) -  v_{f} \boldsymbol{\hat{k}}_{f} \cdot
\boldsymbol{q}} \Bigr\} . \eqa

First, we consider the diagonal component of the boson
self-energy. It is convenient for the momentum integration to take
the angular coordinate \bqa && \Pi_{xx}(\boldsymbol{q},i\Omega)
\nn && = N_{\sigma} \frac{g^{2}}{4v_{f}(2\pi)^{2}} \frac{1}{\beta}
\sum_{i\omega} \int_{0}^{2\pi} d \theta \int_{-\infty}^{\infty} d
\varepsilon \frac{2 \sin^{2} \theta }{i\Omega -
[\Sigma(i\omega+i\Omega) - \Sigma(i\omega)] -
[\Phi(i\omega+i\Omega) - \Phi(i\omega)] - v_{f} q_{x} \cos \theta
- v_{f} q_{y} \sin \theta } \nn && \Bigl\{ \frac{1}{ i\omega -
\Sigma(i\omega) - \Phi(i\omega) - \varepsilon} - \frac{1}{
i\omega+i\Omega - \Sigma(i\omega+i\Omega) - \Phi(i\omega+i\Omega)
- \varepsilon - v_{f} q_{x} \cos \theta - v_{f} q_{y} \sin \theta
} \Bigr\} \nn && = N_{\sigma} \frac{i \pi g^{2}}{4v_{f}(2\pi)^{2}
} \int_{-\infty}^{\infty} \frac{d \omega}{2\pi} \int_{0}^{2\pi} d
\theta \frac{ 2 \sin^{2} \theta [\boldsymbol{sgn} (\omega) -
\boldsymbol{sgn} (\omega+ \Omega)] }{i\Omega -
[\Sigma(i\omega+i\Omega) - \Sigma(i\omega)] -
[\Phi(i\omega+i\Omega) - \Phi(i\omega)] - v_{f} q_{x} \cos \theta
- v_{f} q_{y} \sin \theta } , \eqa where we considered the zero
temperature limit in the last equality.

The angular integration is given by \bqa && \int_{0}^{2\pi} d
\theta \frac{ \sin^{2} \theta }{i \Omega' - v_{f} q_{x} \cos
\theta - v_{f} q_{y} \sin \theta } \nn && = 2 \pi i \Bigl(
\frac{1}{\sqrt{{\Omega'}^{2} + v_{f}^{2} q^{2}}} +
\frac{\Omega'}{v_{f}^{2}q^{2}} +
\frac{{\Omega'}^{2}}{v_{f}^{2}q^{2}} \frac{1}{\sqrt{{\Omega'}^{2}
+ v_{f}^{2} q^{2}}} \Bigr) \frac{q_{x}^{2}}{q^{2}} - 2 \pi i
\Bigl( \frac{\Omega'}{v_{f}^{2}q^{2}} +
\frac{{\Omega'}^{2}}{v_{f}^{2}q^{2}} \frac{1}{\sqrt{{\Omega'}^{2}
+ v_{f}^{2} q^{2}}} \Bigr) \frac{q_{y}^{2}}{q^{2}} , \eqa where
\bqa && \Omega' = \Omega + i [\Sigma(i\omega+i\Omega) -
\Sigma(i\omega)] + i [\Phi(i\omega+i\Omega) - \Phi(i\omega)]
\nonumber \eqa is an effective frequency with both normal and
anomalous self-energies.

Using this angular integration, we find that both $z = 3$ and $z =
2$ critical dynamics appear in the following way \bqa &&
\Pi_{xx}(\boldsymbol{q},i\Omega) \approx N_{\sigma} \frac{
g^{2}}{4v_{f} } \int_{-\infty}^{\infty} \frac{d \omega}{2\pi}
[\boldsymbol{sgn} (\omega) - \boldsymbol{sgn} (\omega+ \Omega)]
\nn && \Bigl\{ \Bigl( \frac{1}{\sqrt{{\Omega}^{2} + v_{f}^{2}
q^{2}}} + \frac{\Omega}{v_{f}^{2}q^{2}} +
\frac{{\Omega}^{2}}{v_{f}^{2}q^{2}} \frac{1}{\sqrt{{\Omega}^{2} +
v_{f}^{2} q^{2}}} \Bigr) \frac{q_{x}^{2}}{q^{2}} - \Bigl(
\frac{\Omega}{v_{f}^{2}q^{2}} +
\frac{{\Omega}^{2}}{v_{f}^{2}q^{2}} \frac{1}{\sqrt{{\Omega}^{2} +
v_{f}^{2} q^{2}}} \Bigr) \frac{q_{y}^{2}}{q^{2}} \Bigr\} \nn &&
\approx N_{\sigma} \frac{g^{2}}{8 \pi v_{f}} \Bigl( \frac{
|\Omega|}{ v_{f} q} + \frac{\Omega|\Omega|}{v_{f}^{2}q^{2}} +
\frac{\Omega^{2}|\Omega|}{v_{f}^{3}q^{3}} \Bigr)
\frac{q_{x}^{2}}{q^{2}} - N_{\sigma} \frac{ g^{2}}{8 \pi v_{f}}
\Bigl( \frac{\Omega|\Omega|}{v_{f}^{2}q^{2}} +
\frac{\Omega^{2}|\Omega|}{v_{f}^{3}q^{3}} \Bigr)
\frac{q_{y}^{2}}{q^{2}} . \eqa A key point in our derivation is
$\Omega' \approx \Omega$ in the low energy limit. In other words,
the normal self-energy is cancelled by the anomalous self-energy
exactly in the low energy limit, explicitly shown in appendix B.
If we do not introduce the anomalous self-energy correction, the
normal self-energy dominates over the bare frequency, changing the
dynamical exponent from $z = 2$ to $z = 12/5$ for the transverse
mode, consistent with the perturbative analysis in nematic quantum
criticality \cite{Garst_Namatic_QCP}.

The other diagonal boson self-energy can be obtained in the
similar way \bqa && \Pi_{yy}(\boldsymbol{q},i\Omega) \approx
N_{\sigma} \frac{g^{2}}{8 \pi v_{f}} \Bigl( \frac{ |\Omega|}{
v_{f} q} + \frac{\Omega|\Omega|}{v_{f}^{2}q^{2}} +
\frac{\Omega^{2}|\Omega|}{v_{f}^{3}q^{3}} \Bigr)
\frac{q_{y}^{2}}{q^{2}} - N_{\sigma} \frac{ g^{2}}{8 \pi v_{f}}
\Bigl( \frac{\Omega|\Omega|}{v_{f}^{2}q^{2}} +
\frac{\Omega^{2}|\Omega|}{v_{f}^{3}q^{3}} \Bigr)
\frac{q_{x}^{2}}{q^{2}} . \eqa

Rewriting the off-diagonal boson self-energy in the angular
coordinate \bqa && \Pi_{xy}(\boldsymbol{q},i\Omega) = - N_{\sigma}
\frac{g^{2}}{4 (2\pi)^{2}} \int_{-\infty}^{\infty} \frac{d
\omega}{2\pi} \int_{0}^{2\pi} d \theta \int_{-\infty}^{\infty} d
\varepsilon \nn && \frac{2 \cos \theta \sin \theta }{i\Omega -
[\Sigma(i\omega+i\Omega) - \Sigma(i\omega)] -
[\Phi(i\omega+i\Omega) - \Phi(i\omega)] - v_{f} q_{x} \cos \theta
- v_{f} q_{y} \sin \theta } \nn && \Bigl\{ \frac{1}{ i\omega -
\Sigma(i\omega) - \Phi(i\omega) - \varepsilon} - \frac{1}{
i\omega+i\Omega - \Sigma(i\omega+i\Omega) - \Phi(i\omega+i\Omega)
- \varepsilon - v_{f} q_{x} \cos \theta - v_{f} q_{y} \sin \theta
} \Bigr\} , \eqa we reach the final expression \bqa &&
\Pi_{xy}(\boldsymbol{q},i\Omega) = N_{\sigma} \frac{ g^{2}}{8 \pi
v_{f} } \Bigl( \frac{|\Omega| }{v_{f}q} + 2
\frac{\Omega|\Omega|}{v_{f}^{2}q^{2}} + 2
\frac{\Omega^{2}|\Omega|}{v_{f}^{3}q^{3} } \Bigr)
\frac{q_{x}q_{y}}{q^{2}} . \eqa

It is straightforward to rewrite Eqs. (A11), (A12), and (A14) with
an introduction of the longitudinal projection operator
$\boldsymbol{P}_{ij}$. Then, we find the $z = 3$ critical dynamics
for the longitudinal mode and the $z = 2$ critical dynamics for
the transverse mode, given by Eq. (7).

\section{Electron self-energy}

In appendix B we derive an electron self-energy. In particular, we
show that the ansatz of Eq. (6) allows the fully self-consistent
Eliashberg solution, where the off-diagonal electron self-energy
turns out to be the same as the normal self-energy. Inserting the
electron Green's function into the Eliashberg equation for the
electron self-energy, we obtain \bqa &&
\boldsymbol{\Sigma}(\boldsymbol{k},i\omega) = g^{2}
\frac{1}{\beta} \sum_{i\Omega} \int \frac{d^{2}
\boldsymbol{q}}{(2\pi)^{2}}
\boldsymbol{D}_{ij}(\boldsymbol{q},i\Omega)
\boldsymbol{\sigma}_{i} \frac{[i\omega+i\Omega + \mu -
\Sigma(i\omega+i\Omega)] \boldsymbol{I} + [v_{f} |\boldsymbol{k} +
\boldsymbol{q}| + \Phi(i\omega+i\Omega) ] \epsilon_{nm}
(\boldsymbol{\hat{k}}_{n} + \boldsymbol{\hat{q}}_{n} )
\boldsymbol{\sigma}_{m}}{[ i\omega+i\Omega + \mu -
\Sigma(i\omega+i\Omega)]^{2} - [v_{f} |\boldsymbol{k} +
\boldsymbol{q}| + \Phi(i\omega+i\Omega) ]^{2} }
\boldsymbol{\sigma}_{j} . \nn  \eqa

Considering Pauli matrix identities Eq. (A3) and taking the
decomposition for the denominator, we can separate the electron
self-energy matrix into two parts \bqa &&
\boldsymbol{\Sigma}(\boldsymbol{k},i\omega) =
\boldsymbol{\Sigma}_{1}(\boldsymbol{k},i\omega) +
\boldsymbol{\Sigma}_{2}(\boldsymbol{k},i\omega) , \eqa where each
part is given by \bqa &&
\boldsymbol{\Sigma}_{1}(\boldsymbol{k},i\omega) = \frac{g^{2}}{2}
\frac{1}{\beta} \sum_{i\Omega} \int \frac{d^{2}
\boldsymbol{q}}{(2\pi)^{2}} \Bigl( D_{L}(\boldsymbol{q},i\Omega) +
D_{T}(\boldsymbol{q},i\Omega) \Bigr) \nn && \Bigl\{ \frac{1}{
i\omega+i\Omega + \mu - \Sigma(i\omega+i\Omega) - v_{f}
|\boldsymbol{k} + \boldsymbol{q}| - \Phi(i\omega+i\Omega) } +
\frac{1}{ i\omega+i\Omega + \mu - \Sigma(i\omega+i\Omega) + v_{f}
|\boldsymbol{k} + \boldsymbol{q}| + \Phi(i\omega+i\Omega) }
\Bigr\} \boldsymbol{I} \nn && - \frac{g^{2}}{2} \frac{1}{\beta}
\sum_{i\Omega} \int \frac{d^{2} \boldsymbol{q}}{(2\pi)^{2}}
\frac{v_{f} \Bigl( (k_{x} + q_{x}) \boldsymbol{\sigma}_{y} -
(k_{y} + q_{y}) \boldsymbol{\sigma}_{x} \Bigr) }{v_{f}
|\boldsymbol{k} + \boldsymbol{q}|} \Bigl(
D_{L}(\boldsymbol{q},i\Omega) + D_{T}(\boldsymbol{q},i\Omega)
\Bigr) \nn && \Bigl\{ \frac{1}{ i\omega+i\Omega + \mu -
\Sigma(i\omega+i\Omega) - v_{f} |\boldsymbol{k} + \boldsymbol{q}|
- \Phi(i\omega+i\Omega) } - \frac{1}{ i\omega+i\Omega + \mu -
\Sigma(i\omega+i\Omega) + v_{f} |\boldsymbol{k} + \boldsymbol{q}|
+ \Phi(i\omega+i\Omega) } \Bigr\}    , \nn &&
\boldsymbol{\Sigma}_{2}(\boldsymbol{k},i\omega) = g^{2}
\frac{1}{\beta} \sum_{i\Omega} \int \frac{d^{2}
\boldsymbol{q}}{(2\pi)^{2}} \Bigl\{ - \Bigl(
D_{L}(\boldsymbol{q},i\Omega)\frac{q_{x}^{2}}{q^{2}} +
D_{T}(\boldsymbol{q},i\Omega) \frac{q_{y}^{2}}{q^{2}} \Bigr)
\frac{k_{y} + q_{y}}{|\boldsymbol{k} + \boldsymbol{q}|}
\boldsymbol{\sigma}_{x} \nn && + \Bigl(
D_{L}(\boldsymbol{q},i\Omega)\frac{q_{y}^{2}}{q^{2}} +
D_{T}(\boldsymbol{q},i\Omega) \frac{q_{x}^{2}}{q^{2}} \Bigr)
\frac{k_{x} + q_{x}}{|\boldsymbol{k} + \boldsymbol{q}|}
\boldsymbol{\sigma}_{y} + \Bigl( [D_{L}(\boldsymbol{q},i\Omega) -
D_{T}(\boldsymbol{q},i\Omega)] \frac{q_{x}q_{y}}{q^{2}} \Bigr)
\Bigl( \frac{k_{x} + q_{x}}{|\boldsymbol{k} + \boldsymbol{q}|}
\boldsymbol{\sigma}_{x} - \frac{k_{y} + q_{y}}{|\boldsymbol{k} +
\boldsymbol{q}|} \boldsymbol{\sigma}_{y} \Bigr) \Bigr\} \nn &&
\Bigl\{ \frac{1}{ i\omega+i\Omega + \mu - \Sigma(i\omega+i\Omega)
- v_{f} |\boldsymbol{k} + \boldsymbol{q}| - \Phi(i\omega+i\Omega)
} - \frac{1}{ i\omega+i\Omega + \mu - \Sigma(i\omega+i\Omega) +
v_{f} |\boldsymbol{k} + \boldsymbol{q}| + \Phi(i\omega+i\Omega) }
\Bigr\} , \eqa respectively. We will see that
$\boldsymbol{\Sigma}_{2}(\boldsymbol{k},i\omega)$ is irrelevant at
low energies, compared with
$\boldsymbol{\Sigma}_{1}(\boldsymbol{k},i\omega)$.

First, we consider
$\boldsymbol{\Sigma}_{1}(\boldsymbol{k},i\omega)$. Performing the
linearization near the chemical potential [Eq. (A6)] and taking
only dominant contributions near the Fermi surface, we obtain the
following expression \bqa &&
\boldsymbol{\Sigma}_{1}(\boldsymbol{k},i\omega) \approx
\frac{g^{2}}{2} \frac{1}{\beta} \sum_{i\Omega} \int \frac{d^{2}
\boldsymbol{q}}{(2\pi)^{2}} \frac{ D_{L}(\boldsymbol{q},i\Omega) +
D_{T}(\boldsymbol{q},i\Omega) }{ i\omega+i\Omega -
\Sigma(i\omega+i\Omega) - \Phi(i\omega+i\Omega) - v_{f}
\boldsymbol{\hat{k}}_{f} \cdot (\boldsymbol{k} -
\boldsymbol{k}_{f}) -  v_{f} \boldsymbol{\hat{k}}_{f} \cdot
\boldsymbol{q}} \boldsymbol{I} \nn && - \frac{g^{2}}{2}
\frac{1}{\beta} \sum_{i\Omega} \int \frac{d^{2}
\boldsymbol{q}}{(2\pi)^{2}} \Bigl( \hat{k}_{x}
\boldsymbol{\sigma}_{y} - \hat{k}_{y} \boldsymbol{\sigma}_{x}
\Bigr) \frac{ D_{L}(\boldsymbol{q},i\Omega) +
D_{T}(\boldsymbol{q},i\Omega) }{ i\omega+i\Omega -
\Sigma(i\omega+i\Omega) - \Phi(i\omega+i\Omega) - v_{f}
\boldsymbol{\hat{k}}_{f} \cdot (\boldsymbol{k} -
\boldsymbol{k}_{f}) -  v_{f} \boldsymbol{\hat{k}}_{f} \cdot
\boldsymbol{q} }  \nn && - \frac{g^{2}}{2} \frac{1}{\beta}
\sum_{i\Omega} \int \frac{d^{2} \boldsymbol{q}}{(2\pi)^{2}} \Bigl(
\frac{q_{x}}{k_{F}} \boldsymbol{\sigma}_{y} - \frac{q_{y}}{k_{F}}
\boldsymbol{\sigma}_{x} \Bigr) \frac{
D_{L}(\boldsymbol{q},i\Omega) + D_{T}(\boldsymbol{q},i\Omega) }{
i\omega+i\Omega - \Sigma(i\omega+i\Omega) - \Phi(i\omega+i\Omega)
- v_{f} \boldsymbol{\hat{k}}_{f} \cdot (\boldsymbol{k} -
\boldsymbol{k}_{f}) -  v_{f} \boldsymbol{\hat{k}}_{f} \cdot
\boldsymbol{q} } .  \eqa

Introducing the polar coordinate \bqa && \boldsymbol{q} = q ( \cos
\theta \boldsymbol{\hat{x}} + \sin \theta \boldsymbol{\hat{y}}) ,
\nn && \boldsymbol{\hat{k}}_{f} = \cos \phi \boldsymbol{\hat{x}} +
\sin \phi \boldsymbol{\hat{y}} , \eqa we can rewrite
$\boldsymbol{\Sigma}_{1}(\boldsymbol{k},i\omega)$ as
$\boldsymbol{\Sigma}_{1}(\boldsymbol{k},i\omega) \approx
\Sigma(i\omega) \boldsymbol{I} + \Phi(i\omega) \epsilon_{ij}
\boldsymbol{\hat{k}}_{i} \boldsymbol{\sigma}_{j} + \Delta
\boldsymbol{\Sigma}_{1}(i\omega,\boldsymbol{k}_{F})$, where \bqa
&& \Delta \boldsymbol{\Sigma}_{1}(i\omega,\boldsymbol{k}_{F}) = -
\frac{g^{2}}{2} \frac{1}{\beta} \sum_{i\Omega} \int_{0}^{\infty} d
q q \int_{0}^{2\pi} d \theta \Bigl( \frac{q}{k_{F}} \cos
(\theta+\phi) \boldsymbol{\sigma}_{y} - \frac{q}{k_{F}} \sin
(\theta+\phi) \boldsymbol{\sigma}_{x} \Bigr) \nn && \frac{
D_{L}(\boldsymbol{q},i\Omega) + D_{T}(\boldsymbol{q},i\Omega) }{
i\omega+i\Omega - \Sigma(i\omega+i\Omega) - \Phi(i\omega+i\Omega)
- v_{f}  q \cos \theta } \eqa can be regarded as a correction for
the Eliashberg self-energies, $\Sigma(i\omega)$ and
$\Phi(i\omega)$ given by Eq. (9).

It is straightforward to perform the angular integral, reaching
the following expression \bqa && \Delta
\boldsymbol{\Sigma}_{1}(i\omega,\boldsymbol{k}_{F}) =
\frac{g^{2}}{2v_{f} k_{F}} \frac{1}{\beta} \sum_{i\Omega}
\int_{0}^{\infty} d q q \frac{q}{k_{F}} \Bigl(
D_{L}(\boldsymbol{q},i\Omega) + D_{T}(\boldsymbol{q},i\Omega)
\Bigr) (\hat{k}_{x} \boldsymbol{\sigma}_{y} - \hat{k}_{y}
\boldsymbol{\sigma}_{x}) \nn && - \frac{g^{2}}{2v_{f} k_{F}}
\frac{1}{\beta} \sum_{i\Omega} \int_{0}^{\infty} d q q
\frac{q}{k_{F}} \Bigl( D_{L}(\boldsymbol{q},i\Omega) +
D_{T}(\boldsymbol{q},i\Omega) \Bigr) \frac{ [\omega+\Omega + i
\Sigma(i\omega+i\Omega) + i \Phi(i\omega+i\Omega)]
\mbox{sgn}(\omega+\Omega) }{ \sqrt{[\omega+\Omega + i
\Sigma(i\omega+i\Omega) + i \Phi(i\omega+i\Omega)]^{2} +
(v_{f}q)^{2}} } (\hat{k}_{x} \boldsymbol{\sigma}_{y} - \hat{k}_{y}
\boldsymbol{\sigma}_{x}) \nn && - \frac{g^{2}}{2v_{f} k_{F}}
\frac{1}{\beta} \sum_{i\Omega} \int_{0}^{\infty} d q q
\frac{q}{k_{F}} \Bigl( D_{L}(\boldsymbol{q},i\Omega) +
D_{T}(\boldsymbol{q},i\Omega) \Bigr) \ln \Bigl(
\frac{i\omega+i\Omega - \Sigma(i\omega+i\Omega) -
\Phi(i\omega+i\Omega) - v_{f}q}{i\omega+i\Omega -
\Sigma(i\omega+i\Omega) - \Phi(i\omega+i\Omega) + v_{f}q} \Bigr) (
\hat{k}_{x} \boldsymbol{\sigma}_{x} + \hat{k}_{y}
\boldsymbol{\sigma}_{y}).  \nn \eqa Comparing this expression with
the Eliashberg self-energies, we find that $\Delta
\boldsymbol{\Sigma}_{1}(i\omega,\boldsymbol{k}_{F})$ is irrelevant
at low energies due to the fact that the scaling dimension of the
argument itself in the integral expression is higher than that of
the Eliashberg solution in addition to the additional momentum
factor $q/k_{F}$. As a result, we obtain \bqa &&
\boldsymbol{\Sigma}_{1} (i\omega,\boldsymbol{k}_{F}) =
\Sigma(i\omega) \boldsymbol{I} + \Phi(i\omega) ( \hat{k}_{x}
\boldsymbol{\sigma}_{y} - \hat{k}_{y} \boldsymbol{\sigma}_{x} ) ,
\nn && \Sigma(i\omega) = - \Phi(i\omega) = - i \frac{g^{2}}{2}
\frac{1}{\beta} \sum_{i\Omega} \int_{0}^{\infty} d q q  \Bigl(
D_{L}(\boldsymbol{q},i\Omega) + D_{T}(\boldsymbol{q},i\Omega)
\Bigr) \frac{ \mbox{sgn}(\omega+\Omega) }{ \sqrt{[\omega+\Omega +
i \Sigma(i\omega+i\Omega) + i \Phi(i\omega+i\Omega)]^{2} + v_{f}
q^{2}}}
%
%
. \nn \eqa

Next, we show that
$\boldsymbol{\Sigma}_{2}(i\omega,\boldsymbol{k}_{F})$ is
irrelevant at low energies. Performing the linearization near the
chemical potential [Eq. (A6)] and taking only dominant
contributions near the Fermi surface, we obtain the following
expression \bqa &&
\boldsymbol{\Sigma}_{2}(i\omega,\boldsymbol{k}_{F}) \approx g^{2}
\frac{1}{\beta} \sum_{i\Omega} \int \frac{d^{2}
\boldsymbol{q}}{(2\pi)^{2}} \Bigl\{ - \Bigl(
D_{L}(\boldsymbol{q},i\Omega) \frac{q_{x}^{2}}{q^{2}} +
D_{T}(\boldsymbol{q},i\Omega) \frac{q_{y}^{2}}{q^{2}} \Bigr)
\hat{\boldsymbol{k}}_{y} \boldsymbol{\sigma}_{x} + \Bigl(
D_{L}(\boldsymbol{q},i\Omega) \frac{q_{y}^{2}}{q^{2}} +
D_{T}(\boldsymbol{q},i\Omega) \frac{q_{x}^{2}}{q^{2}} \Bigr)
\hat{\boldsymbol{k}}_{x} \boldsymbol{\sigma}_{y} \Bigr\} \nn &&
\frac{ 1 }{ i\omega+i\Omega - \Sigma(i\omega+i\Omega) -
\Phi(i\omega+i\Omega) - v_{f} \boldsymbol{\hat{k}}_{f} \cdot
\boldsymbol{q} } \nn && + g^{2} \frac{1}{\beta} \sum_{i\Omega}
\int \frac{d^{2} \boldsymbol{q}}{(2\pi)^{2}} \Bigl\{ - \Bigl(
D_{L}(\boldsymbol{q},i\Omega) \frac{q_{x}^{2}}{q^{2}} +
D_{T}(\boldsymbol{q},i\Omega) \frac{q_{y}^{2}}{q^{2}} \Bigr)
\frac{\boldsymbol{q}_{y}}{k_{F}} \boldsymbol{\sigma}_{x} + \Bigl(
D_{L}(\boldsymbol{q},i\Omega) \frac{q_{y}^{2}}{q^{2}} +
D_{T}(\boldsymbol{q},i\Omega) \frac{q_{x}^{2}}{q^{2}} \Bigr)
\frac{\boldsymbol{q}_{x}}{k_{F}} \boldsymbol{\sigma}_{y} \Bigr\}
\nn && \frac{ 1 }{i\omega+i\Omega - \Sigma(i\omega+i\Omega) -
\Phi(i\omega+i\Omega) - v_{f} \boldsymbol{\hat{k}}_{f} \cdot
\boldsymbol{q} } \nn && + g^{2} \frac{1}{\beta} \sum_{i\Omega}
\int \frac{d^{2} \boldsymbol{q}}{(2\pi)^{2}} \Bigl(
[D_{L}(\boldsymbol{q},i\Omega) - D_{T}(\boldsymbol{q},i\Omega)]
\frac{q_{x}q_{y}}{q^{2}} \Bigr)  \Bigl( \hat{\boldsymbol{k}}_{x}
\boldsymbol{\sigma}_{x} - \hat{\boldsymbol{k}}_{y}
\boldsymbol{\sigma}_{y} \Bigr) \frac{ 1 }{ i\omega+i\Omega -
\Sigma(i\omega+i\Omega) - \Phi(i\omega+i\Omega) - v_{f}
\boldsymbol{\hat{k}}_{f} \cdot \boldsymbol{q} }\nn && + g^{2}
\frac{1}{\beta} \sum_{i\Omega} \int \frac{d^{2}
\boldsymbol{q}}{(2\pi)^{2}} \Bigl( [D_{L}(\boldsymbol{q},i\Omega)
- D_{T}(\boldsymbol{q},i\Omega)] \frac{q_{x}q_{y}}{q^{2}} \Bigr)
\Bigl( \frac{\boldsymbol{q}_{x}}{k_{F}} \boldsymbol{\sigma}_{x} -
\frac{\boldsymbol{q}_{y}}{k_{F}} \boldsymbol{\sigma}_{y} \Bigr)
\frac{ 1 }{ i\omega+i\Omega - \Sigma(i\omega+i\Omega) -
\Phi(i\omega+i\Omega) -  v_{f} \boldsymbol{\hat{k}}_{f} \cdot
\boldsymbol{q} } . \nn \eqa

Introducing the polar coordinate of Eq. (B5), we rewrite the above
expression as follows \bqa &&
\boldsymbol{\Sigma}_{2}(i\omega,\boldsymbol{k}_{F}) =
\frac{g^{2}}{4\pi^{2}} \frac{1}{\beta} \sum_{i\Omega}
\int_{0}^{\infty} d q q \int_{0}^{2\pi} d \theta \Bigl\{ - \Bigl(
D_{L}(\boldsymbol{q},i\Omega) \cos^{2} \theta +
D_{T}(\boldsymbol{q},i\Omega) \sin^{2} \theta \Bigr) \sin \phi
\boldsymbol{\sigma}_{x} \nn && + \Bigl(
D_{L}(\boldsymbol{q},i\Omega) \sin^{2} \theta +
D_{T}(\boldsymbol{q},i\Omega) \cos^{2} \theta \Bigr) cos\phi
\boldsymbol{\sigma}_{y} \Bigr\} \frac{ 1 }{ i\omega+i\Omega -
\Sigma(i\omega+i\Omega) - \Phi(i\omega+i\Omega) - v_{f} \cos
(\theta-\phi) } \nn && + \frac{g^{2}}{4\pi^{2}} \frac{1}{\beta}
\sum_{i\Omega} \int_{0}^{\infty} d q q \int_{0}^{2\pi} d \theta
\Bigl\{ - \Bigl( D_{L}(\boldsymbol{q},i\Omega) \cos^{2}\theta +
D_{T}(\boldsymbol{q},i\Omega) \sin^{2}\theta \Bigr)
\frac{q}{k_{F}} \sin\theta \boldsymbol{\sigma}_{x} \nn && + \Bigl(
D_{L}(\boldsymbol{q},i\Omega) \sin^{2}\theta +
D_{T}(\boldsymbol{q},i\Omega) \cos^{2}\theta \Bigr)
\frac{q}{k_{F}} \cos\theta \boldsymbol{\sigma}_{y} \Bigr\} \frac{
1 }{ i\omega+i\Omega - \Sigma(i\omega+i\Omega) -
\Phi(i\omega+i\Omega) - v_{f} \cos (\theta-\phi) } \nn && +
\frac{g^{2}}{4\pi^{2}} \frac{1}{\beta} \sum_{i\Omega}
\int_{0}^{\infty} d q q \int_{0}^{2\pi} d \theta
[D_{L}(\boldsymbol{q},i\Omega) - D_{T}(\boldsymbol{q},i\Omega)]
\Bigl( \cos\phi \boldsymbol{\sigma}_{x} - \sin\phi
\boldsymbol{\sigma}_{y} \Bigr) \nn && \frac{ \cos\theta\sin\theta
}{ i\omega+i\Omega - \Sigma(i\omega+i\Omega) -
\Phi(i\omega+i\Omega) - v_{f} \cos (\theta-\phi) } \nn && +
\frac{g^{2}}{4\pi^{2}} \frac{1}{\beta} \sum_{i\Omega}
\int_{0}^{\infty} d q q \int_{0}^{2\pi} d \theta
[D_{L}(\boldsymbol{q},i\Omega) - D_{T}(\boldsymbol{q},i\Omega)]
\Bigr)  \Bigl( \frac{q}{k_{F}} \cos\theta \boldsymbol{\sigma}_{x}
- \frac{q}{k_{F}} \sin\theta \boldsymbol{\sigma}_{y} \Bigr) \nn &&
\frac{ \cos\theta\sin\theta }{ i\omega+i\Omega -
\Sigma(i\omega+i\Omega) - \Phi(i\omega+i\Omega) - v_{f} \cos
(\theta-\phi) } . \eqa Performing the angular integration, one can
see irrelevance of this contribution, basically resulting from
higher order momentum integrals. We prove that our ansatz is fully
self-consistent in the Nambu-Eliashberg approximation.

\end{widetext}


\begin{thebibliography}{9}
\bibitem{Review_Berry} D. Xiao, M.-C. Chang, and Q. Niu, Rev. Mod. Phys. {\bf 82},
1959 (2010).
\bibitem{Review_AHE} N. Nagaosa, J. Sinova, S. Onoda, A. H. MacDonald, and N. P.
Ong, Rev. Mod. Phys. {\bf 82}, 1539 (2010).
\bibitem{Review_TI} X.-L. Qi and S.-C. Zhang, arXiv:1008.2026
(unpublished).
\bibitem{Soliton_TextBook} R. Rajaraman, \textit{Solitons and
Instantons}(Elsevier Science, New York, 2003).
\bibitem{Review_QCP} P. Gegenwart, Q. Si, and F. Steglich,
Nature Physics {\bf 4}, 186 (2008); H. v. Lohneysen, A. Rosch, M.
Vojta, and P. Wolfle, Rev. Mod. Phys. {\bf 79}, 1015 (2007).
\bibitem{Xu_RPA} C. Xu, Phys. Rev. B {\bf 81}, 054403 (2010).
\bibitem{Rech_FMQCP} J. Rech, C. Pepin, and A. V. Chubukov, Phys. Rev.
B {\bf 74}, 195126 (2006).
\bibitem{Kim_LW} A. Benlagra, K.-S. Kim, C. P\'epin,
arXiv:0902.3630 (unpublished).
\bibitem{Garst_Namatic_QCP} M. Zacharias, P. Wolfle, and M. Garst,
Phys. Rev. B {\bf 80}, 165116 (2009).
\bibitem{Shindou_E} R. Shindou and L. Balents, Phys. Rev. Lett. {\bf 97}, 216601
(2006).
\bibitem{FM_RKKY} Q. Liu, C.-X. Liu, C. Xu, X.-L. Qi,
and S.-C. Zhang, Phys. Rev. Lett. {\bf 102}, 156603 (2009).
\bibitem{Tran_Kim_MI_TI} Minh-Tien Tran and Ki-Seok Kim, Phys. Rev. B {\bf 82}, 155142
(2010).
\bibitem{Ryu_Dirac_Mass} S. Ryu, C. Mudry, C.-Y. Hou, and C.
Chamon, Phys. Rev. B {\bf 80}, 205319 (2009).
\bibitem{Graphene_QMC} Z. Y. Meng, T. C. Lang, S. Wessel, F. F. Assaad, and
A. Muramatsu, Nature {\bf 464}, 847 (2010).
\bibitem{Graphene_Interaction} F. Wang, Phys. Rev. B {\bf 82}, 024419
(2010); Y.-M. Lu and Y. Ran, arXiv:1007.3266 (unpublished); B. K.
Clark, D. A. Abanin, and S. L. Sondhi, arXiv:1010.3011
(unpublished); G. Wang, M. O. Goerbig, B. Gremaud, and C.
Miniatura, arXiv:1006.4456 (unpublished); A. Vaezi and X.-G. Wen,
arXiv:1010.5744 (unpublished);  Minh-Tien Tran and Ki-Seok Kim,
arXiv:1011.1700 (unpublished).
\bibitem{LW} J. M. Luttinger and J. C. Ward, Phys. Rev. {\bf 118}, 1417
(1960); G. Baym and L. P. Kadanoff, Phys. Rev. {\bf 124}, 287
(1961).
\bibitem{Rech_FMQCP_Comment} We would like to emphasize that we are
considering the case of finite chemical potentials.
\bibitem{Garst_Nematic} M. Garst and A. V. Chubukov, Phys. Rev. B
{\bf 81}, 235105 (2010).
\bibitem{Kim_AL} Ki-Seok Kim, Phys. Rev. B {\bf 83}, 035123 (2011).
\bibitem{SungSik_Genus} Sung-Sik Lee, Phys. Rev. B {\bf 80}, 165102
(2009).
\bibitem{Metlitski_Sachdev1} Max A. Metlitski and S. Sachdev, Phys. Rev. B {\bf 82}, 075127 (2010).
\bibitem{Metlitski_Sachdev2} Max A. Metlitski and S. Sachdev, Phys. Rev. B {\bf 82}, 075128 (2010).
\bibitem{McGreevy} David F. Mross, J. McGreevy, H. Liu, and T.
Senthil, Phys. Rev. B {\bf 82}, 045121 (2010).
\bibitem{Kim_Ladder} Ki-Seok Kim, Phys. Rev. B {\bf 82}, 075129
(2010).
\bibitem{HFQCP} P. Gegenwart, Q. Si, and F. Steglich,
Nature Physics {\bf 4}, 186 (2008); H. v. Lohneysen, A. Rosch, M.
Vojta, and P. Wolfle, Rev. Mod. Phys. {\bf 79}, 1015 (2007).
\bibitem{Kim_Adel_Pepin} K.-S. Kim, A. Benlagra, and C. P\'epin,
Phys. Rev. Lett. {\bf 101}, 246403 (2008).
\bibitem{Kim_TR} K.-S. Kim and C. P\'epin,
Phys. Rev. Lett. {\bf 102}, 156404 (2009); K.-S. Kim and C.
P\'epin, J. Phys.: Condens. Matter {\bf 22}, 025601 (2010).
\bibitem{Kim_Jia} K.-S. Kim and C. Jia, Phys. Rev. Lett. {\bf 104},
156403 (2010).
\bibitem{Kim_Pepin_Seebeck} K.-S. Kim and C. P\'epin,
arXiv:1005.3354, to be published in Phys. Rev. B; K.-S. Kim and C.
P\'epin, Phys. Rev. B {\bf 81}, 205108 (2010).
\bibitem{BCS_Book} J. R. Schrieffer, \textit{Theory of Superconductivity}
(Westview Press, Colorado, 1999).
\bibitem{Spin_Book} A. Auerbach, \textit{Interacting Electrons and Quantum
magnetism }(Springer-Verlag, New York, 1994).
\end{thebibliography}
\end{document}